\documentclass[%
 reprint,
superscriptaddress,
nofootinbib,
 amsmath,amssymb,
 aps,
]{revtex4-2}
\usepackage{multirow}
\usepackage{subfigure}
\usepackage{graphicx}
\usepackage{dcolumn}
\usepackage{bm}
\usepackage[colorlinks,
            linkcolor=red,
            anchorcolor=blue,
            citecolor=blue
            ]{hyperref}

\graphicspath{ {figure/} }
\begin{document}

\title{Prediction of lithium isotope fluxes using data-driven production cross sections}

\author{Meng-Jie Zhao}
\email{zhaomj@ihep.ac.cn}
 \affiliation{%
 Key Laboratory of Particle Astrophysics, Institute of High Energy Physics, Chinese Academy of Sciences, Beijing 100049, China}
\affiliation{
China Center of Advanced Science and Technology, Beijing 100190, China 
}%
 \author{Xiao-Jun Bi}
 \email{bixj@ihep.ac.cn}
\affiliation{%
 Key Laboratory of Particle Astrophysics, Institute of High Energy Physics, Chinese Academy of Sciences, Beijing 100049, China}
\affiliation{
 University of Chinese Academy of Sciences, Beijing 100049, China 
}%
\author{Kun Fang}
\email{fangkun@ihep.ac.cn}
\affiliation{%
 Key Laboratory of Particle Astrophysics, Institute of High Energy Physics, Chinese Academy of Sciences, Beijing 100049, China}
 \author{Xing-Jian Lv}
\email{lvxj@ihep.ac.cn}
 \affiliation{%
 Key Laboratory of Particle Astrophysics, Institute of High Energy Physics, Chinese Academy of Sciences, Beijing 100049, China}
\affiliation{
 University of Chinese Academy of Sciences, Beijing 100049, China 
}%
 \author{Peng-Fei Yin}
\email{yinpf@ihep.ac.cn}
\affiliation{%
 Key Laboratory of Particle Astrophysics, Institute of High Energy Physics, Chinese Academy of Sciences, Beijing 100049, China}



\date{\today}

\begin{abstract}
Galactic cosmic rays (CRs) generally share common propagation features, leading to consistent spectral observations of secondary nuclei such as Li, Be, and B. However, the Li spectrum predicted by the CR diffusion coefficient inferred from B/C is significantly lower than the latest measurement of AMS-02. 
This anomaly may be attributed to the missing contributions from the heavy nuclei components in cosmic rays.
By including these missing contributions the excess of the Li spectrum disappears. 
However, another inconsistency still exists since the calculated Li spectrum is now overestimated compared to the data. In this work, we update the cross-section model used to calculate the Li production according to more cross-section measurements.
We find that the cross sections of these added reactions are systematically overestimated, and should be renormalized to the interpolations of available data.
As a result, our prediction of the total Li spectrum is consistent with the measurement without discrepancy, and our prediction of the $\rm^6Li$ and $\rm^7Li$ spectra are consistent with the preliminary measurements of AMS-02 within the cross-section uncertainties.
\end{abstract}
\maketitle


\section{\label{sec:level1}INTRODUCTION}
The study of galactic CRs focuses on updating our knowledge of the CR acceleration and propagation mechanism, the spatial distribution of the CR sources and the interstellar medium (ISM), and the inelastic and production cross sections of nuclei reactions.
Based on this knowledge, the standard CR propagation framework \cite{Ginzburg:1976dj, Trotta:2010mx, Genolini:2019ewc, Yuan:2018vgk} was gradually constructed and frequently used to analyze the CR measurements, such as PAMELA \cite{Adriani:2011cu} and AMS-02 \cite{Aguilar:2018njt, Consolandi:2016fhd, Aguilar:2017hno, Aguilar:2021tos, AMS:2021brg}.

Secondary CRs mainly arise from the fragmentation of heavier nuclei upon collision with the ISM gas. Therefore, the flux ratio of secondaries to their primary nuclei informs us about the grammages of CRs\footnote{The integrated gas density along the CR propagation path before they escape from the Galaxy.} as well as the production cross sections of fragments.
The Li, Be, and B have the highest abundance among secondary CRs and have been measured by AMS-02 collaboration \cite{Aguilar:2018njt, Aguilar:2021tos} with impressive precision levels.
As they have similar nuclei charges and are mainly produced by progenitors of C and O, the propagation consistency of these CRs is expected. This means secondary-to-primary ratios such as B/C, B/O, Be/C, and Li/C can be reproduced within the same propagation framework to match the measurements.

The propagation consistency has been studied by many groups~\cite{Boschini:2019gow, Luque:2021nxb, delaTorreLuque:2022vhm, Weinrich:2020cmw, Korsmeier:2021brc}, and the Be and B spectra were well-reproduced above 10 GV assuming the same propagation parameters, with inconsistency of Be/B at several GVs~\cite{Boschini:2019gow, Luque:2021nxb, delaTorreLuque:2022vhm}.
As for the Li spectrum, they found an unexpected excess in the experimental AMS-02 result~\cite{Aguilar:2018njt, Aguilar:2021tos} compared to the prediction of propagation framework, and several interpretations were suggested such as an extra primary $\rm^7Li$ component~\cite{Kawanaka:2017cae, Boschini:2019gow} or the adjustments of production cross section~\cite{Luque:2021nxb, delaTorreLuque:2022vhm, Weinrich:2020cmw, Korsmeier:2021brc}.
A recent work~\cite{Maurin:2022irz} emphasized that most previous interpretations rely on a cross-section set that does not account for the cross sections of Li from heavy progenitors, such as Ne, Mg, Si, and Fe.
By adding the lithium production contributions from missing progenitors, the Li flux can be enhanced by $\sim$ 20-50\%, and the need for an extra Li component disappears.
However, this enhancement would lead to an overproduction of Li if regarding the B spectrum as a reference, according to the cross-section sets of OPT12up22 and OPT22 in Ref.~\cite{Maurin:2022irz}.
Whether it is associated with the misunderstanding of these unmeasured reactions deserves further inspection.

The uncertainty in production cross sections is an essential consideration when using secondary species measurements to investigate CR propagation.
To accurately calculate the production of secondaries, we have updated the cross-section database and the cross-section model according to the available measurements. In Ref.~\cite{Zhao:2024qbj}, we resolved the inconsistency of Be/B at several GVs by improving the cross sections of Be and B, and the predicted Be isotopes were consistent with the preliminary measurements of AMS-02~\cite{AMS_icrc2021, AMS_ichep}. In Ref.~\cite{Lv:2024tls}, the recently measured deuteron spectrum \cite{AMS:2024idr} was evaluated with the updated cross-section model, and we found that a primary deuteron component is needed to explain the excess at high energies.

In this paper, we aim to resolve the overproduction of the Li spectrum found in Ref.~\cite{Maurin:2022irz} by improving the cross-section models.
We add the lithium production contributions from missing progenitors according to Tsao and Silberberg's parametrization~\cite{Silberberg:1998lxa}.
Considering the poorly measured cross sections of Li, the calculation relies seriously on the semi-empirical results of parametrization with large systematical uncertainties.
To ensure reliability, these less-constrained reactions will be estimated and renormalized according to the available measurements.
The forthcoming AMS-02 data of Li isotope spectra would be useful to examine the source of the Li anomaly by separating the isotopes of $\rm^6Li$ and $\rm^7Li$.
Therefore we will list the contribution fractions of different channels individually for $\rm^6Li$ and $\rm^7Li$, and make predictions for the Li isotope spectra as well as analyze the cross-section uncertainties.
There are several differences between Ref.~\cite{Maurin:2022irz} and our work.
They obtained the propagation parameters by fitting the B/C, Be/C, and Li/C ratios with nuisance parameters of Li, Be, and B cross sections. In comparison, the propagation parameters in the paper are based on our previous work~\cite{Zhao:2024qbj}, where B and $\rm^7Be$ data were well-fitted simultaneously with small nuisances of cross sections\footnote{We thought that the cross sections of B and $\rm^7Be$ are well-constrained by measurements, hence the spectra of them are more reliable than Li spectrum.}.
For unmeasured reactions of Li from projectiles heavier than O, they directly used the result of parametrization, while we renormalize them together according to the interpolations of available data.

This paper is organized as follows. 
In Sec.~\ref{sec:level2}, we introduce the CR propagation model and the cross-section setup for analysis. For each reaction channel, we ranked their contributions and estimated their cross-section uncertainties.
In Sec.~\ref{sec:result}, we compare the theoretical predictions of $\rm^6Li$, $\rm^7Li$, and the total Li spectra with the measured data. 
Finally, Sec.~\ref{sec:conclusion} is the summary of our findings above.
For improved readability, we have moved several figures and discussions regarding the updated cross-section data and parameterizations used in this study to the Appendix.

\section{\label{sec:level2}CALCULATION SETUP}
\subsection{Propagation}
We adopt the standard CR propagation model with reacceleration, which is frequently used in CR analysis \cite{Trotta:2010mx, Genolini:2019ewc, Yuan:2018vgk}. Generally, the propagation equation of Galactic CRs is expressed as
\begin{eqnarray}
	 {\frac{\partial \psi}{\partial t}}=&&q(x,p)+\nabla\cdot(D_{xx}\nabla\psi-V_c\psi)
	 +{\frac{\partial}{\partial p}}[p^2D_{pp}{\frac{\partial}{\partial p}}({\frac{\psi}{p^2}})]\nonumber \\\label{eq:trans1}
	 &&-{\frac{\partial}{\partial p}}[\dot p\psi-{\frac{p}{3}}(\nabla\cdot V_c)\psi]
	-{\frac{\psi}{\tau_f}}-{\frac{\psi}{\tau_r}}\,, \label{eq:trans2}
\end{eqnarray}
where $\psi$ is the density per unit of particle momentum, $q(x,p)$ is the source distribution, $D_{xx}$ is the spatial diffusion coefficient, $V_c$ is the convection velocity, $D_{pp}$ is the momentum space diffusion coefficient, $\dot p\equiv dp/dt$ is the ionization and Coulomb losses terms, $\tau_f$ is the time scales for particle fragmentation, and $\tau_r$ is the time scales for radioactive decay. To solve the propagation equation, we adopt the numerical framework of {\footnotesize GALPROP} v57\footnote{The current version is available at \url{https://galprop.stanford.edu}.} \cite{Strong:1998pw, Strong:1998fr, Porter:2021tlr}, which includes a nuclear reaction network to calculate the one-step and multistep fragmentation of particles.
Our data-driven parametrization is based on this reaction network with some improvements introduced in Sec.~\ref{sec:xs} and Appendix.~\ref{app:xsdata}.

In this framework, the diffusion coefficient $D_{xx}$ is parameterized as a broken power-law by
\begin{equation}
\label{eq:diff}
D_{xx}=D_0\beta^\eta\left({\frac{R_{h}}{R_0}}\right)^{\delta_1}\left\{\begin{array}{cc}
\left(R / R_1\right)^{\delta_1} , & R\leq R_h \\
\left(R / R_1\right)^{\delta_2} , & R> R_h\; ,
\end{array}\right.
\end{equation}
where $R_0=4$~GV is the reference rigidity, $R_h$ is the break rigidity, $\beta=v/c$ is the particle velocity divided by the speed of light, and the low-energy random-walk process is shaped by the factor $\beta^\eta$. Here $\eta\ne1$ is introduced to improve the calculated B/C ratio at low rigidity to fit the observations. The harder index $\delta_2$ above the high-energy break rigidity $R_h$ is assumed to fit the hardening of B/C and B/O ratios observed by AMS-02~\cite{Aguilar:2021tos}.

The scattering of CR particles on randomly moving magnetohydrodynamics waves leads to stochastic acceleration, which is described in the transport equation as diffusion in momentum space $D_{pp}$. Considering the scenario where the CRs are reaccelerated by colliding with the interstellar random weak hydrodynamic waves, the relation between the spatial diffusion coefficient $D_{xx}$ and the momentum diffusion coefficient $D_{pp}$ is expressed as \cite{1994ApJ...431..705S}:

\begin{equation}
	D_{xx}D_{pp}=\frac{4p^2V_a^2}{3\delta(4-\delta)(4-\delta^2)\omega}\,.
\end{equation}

\begin{table}
\centering
\caption{The propagation parameters used in this work, taken from Ref.~\cite{Zhao:2024qbj}. \label{table: propagation}}
\begin{tabular}{cccccccc}
\hline
\hline
$D_0$ & $\delta_1$ & $\delta_2$ & $R_h$ & $z_h$ & $V_A$ & $\eta$ & $\phi$\\
($10^{28} \text{cm}^2 \text{s}^{-1}$) & & & (GV) & (kpc) & (km s$^{-1}$) & & (GV)\\
\hline
5.197 & 0.45 & 0.215 & 280 & 5.674 & 17.809 & -0.484 & 0.72 \\
\hline
\end{tabular}
\end{table}

\begin{table}[ht]
\centering
\caption{The injection parameters and source abundances used in this work. The abundance of the proton is fixed to $1.06\times10^6$ at 100 GeV$/n$.}
\begin{tabular}{ccccc}
\hline
\hline
Elements & $\nu_0$ & $R_{br}$(GV) & $\nu_1$ & abundance\\
\hline
$\rm C$ & 1.249 & 2.088 & 2.365 & 3201.0 \\
$\rm N$ & 1.249 & 2.088 & 2.365 & 275.8 \\
$\rm O$ & 1.249 & 2.088 & 2.390 & 4115.5 \\
$\rm Ne$ & 1.1 & 2.374 & 2.36 & 549.0 \\
$\rm Na$ & 1.1 & 2.374 & 2.44 & 22.8 \\
$\rm Mg$ & 1.1 & 2.374 & 2.41 & 804.1 \\
$\rm Al$ & 1.1 & 2.374 & 2.41 & 65.0 \\
$\rm Si$ & 1.1 & 2.374 & 2.41 & 759.7 \\
$\rm S$ & 0.847 & 2.374 & 2.425 & 120.7 \\
$\rm Fe$ & 2.15 & 25.06 & 2.385 & 705.0 \\
\hline
\label{table: injection}
\end{tabular}
\end{table}

\begin{figure}[htbp]
\includegraphics[width=0.45\textwidth,trim=0 0 0 0,clip]{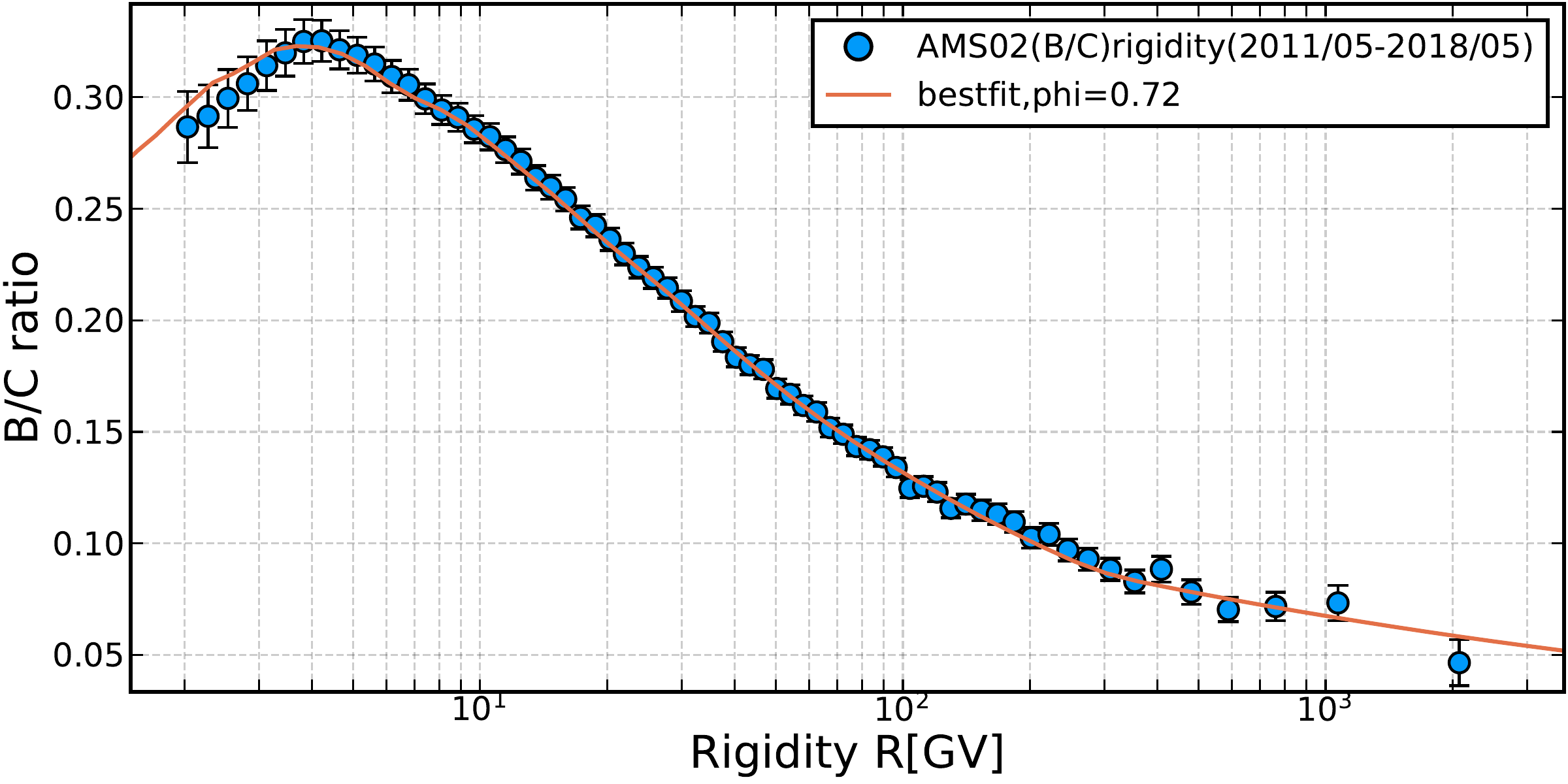}
\includegraphics[width=0.45\textwidth,trim=0 0 0 0,clip]{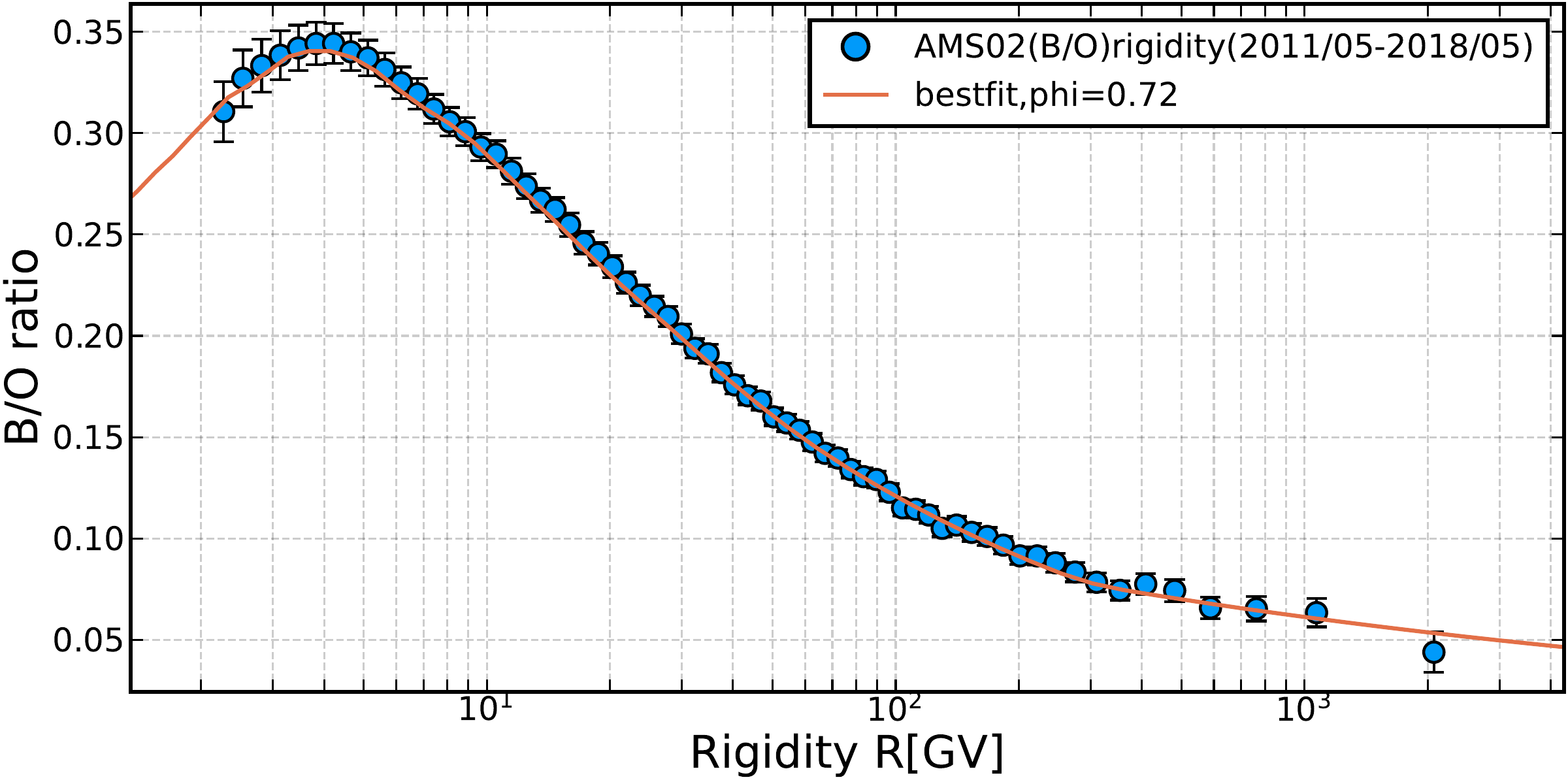}
\caption{\label{fig:ratio} 
B/C and B/O ratio calculated with the best-fit parameters, compared with the experimental data: AMS-02~\cite{Aguilar:2021tos}.
}
\end{figure}

Our previous work \cite{Zhao:2024qbj} analyzed the preliminary Be isotopes and the B spectra based on data-driven production cross sections. As mid-mass secondary particles, B, $\rm^7Be$, and $\rm^{10}Be$ data are well-fitted simultaneously, implying that the constrained parameters can also be used to predict the production of Li isotopes.
In this work, we adopt the diffusion/reacceleration model with almost the same setup of the propagation parameters from Ref.~\cite{Zhao:2024qbj}, listed in Table~\ref{table: propagation}. In Figure.~\ref{fig:ratio}, we illustrate the best-fit B/C and B/O ratios compared with the observations of AMS-02~\cite{Aguilar:2021tos}.
It is important to note that, we have made several updates to the nuclear database and adjusted the parameterizations of dominant channels based on recent observations. 
The B production from C at high energies is enhanced by including the [NA61/SHINE] data~\cite{Amin:2021oow, Amin:2023fki}, and productions from subdominant projectiles of Ne, Mg, and Si are enhanced by including the charge-changing cross-section data provided in Ref.~\cite{Webber:1990kb, Zeitlin:2001ye, Zeitlin:2007sm, Zeitlin:2011qg}.
To fit the B/C ratio, the required diffusion coefficient in Table~\ref{table: propagation} should be larger than that obtained with old cross-section models.

\begin{figure*}[htbp]
\includegraphics[width=0.45\textwidth,trim=0 0 0 0,clip]{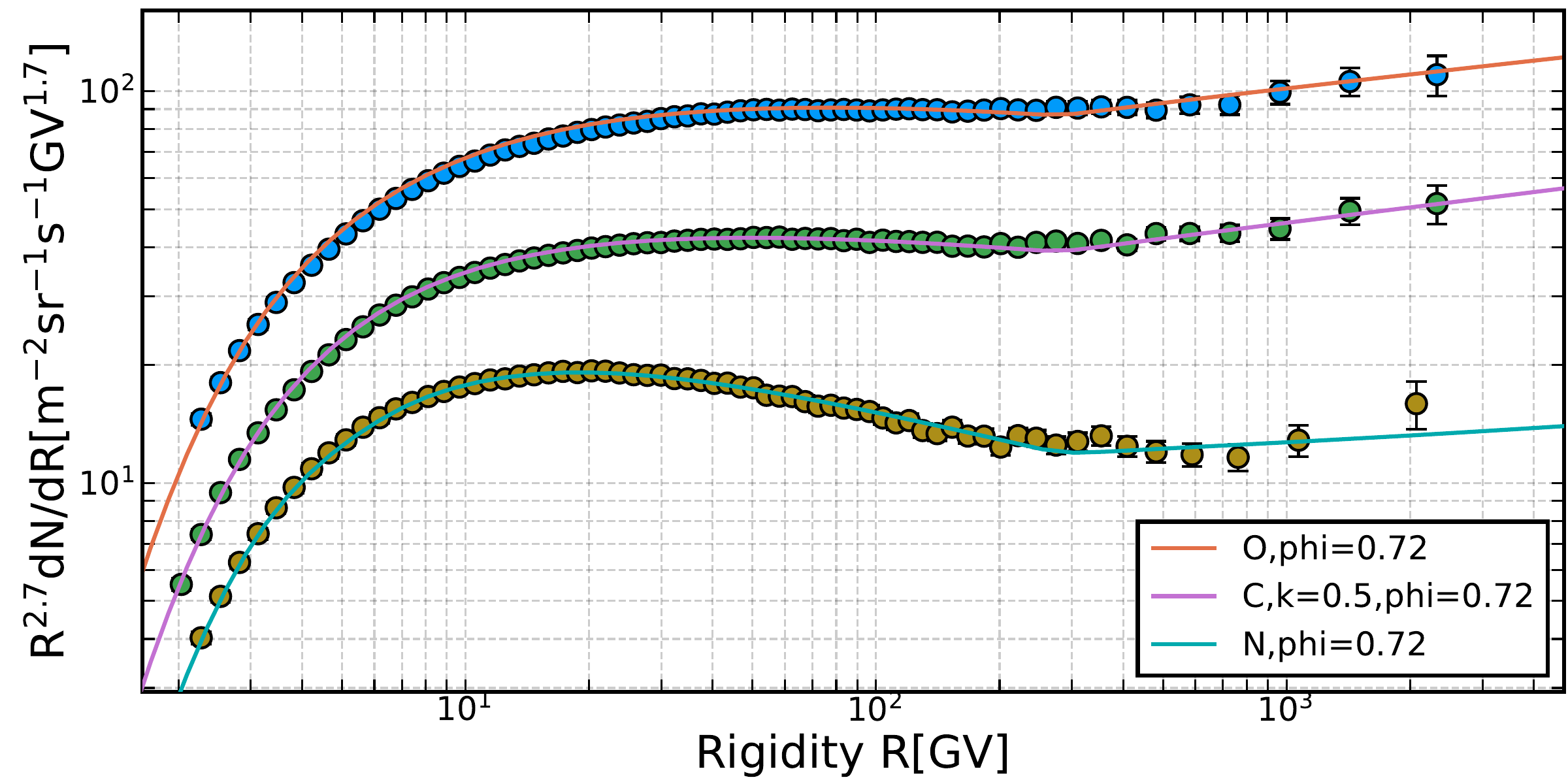}
\includegraphics[width=0.45\textwidth,trim=0 0 0 0,clip]{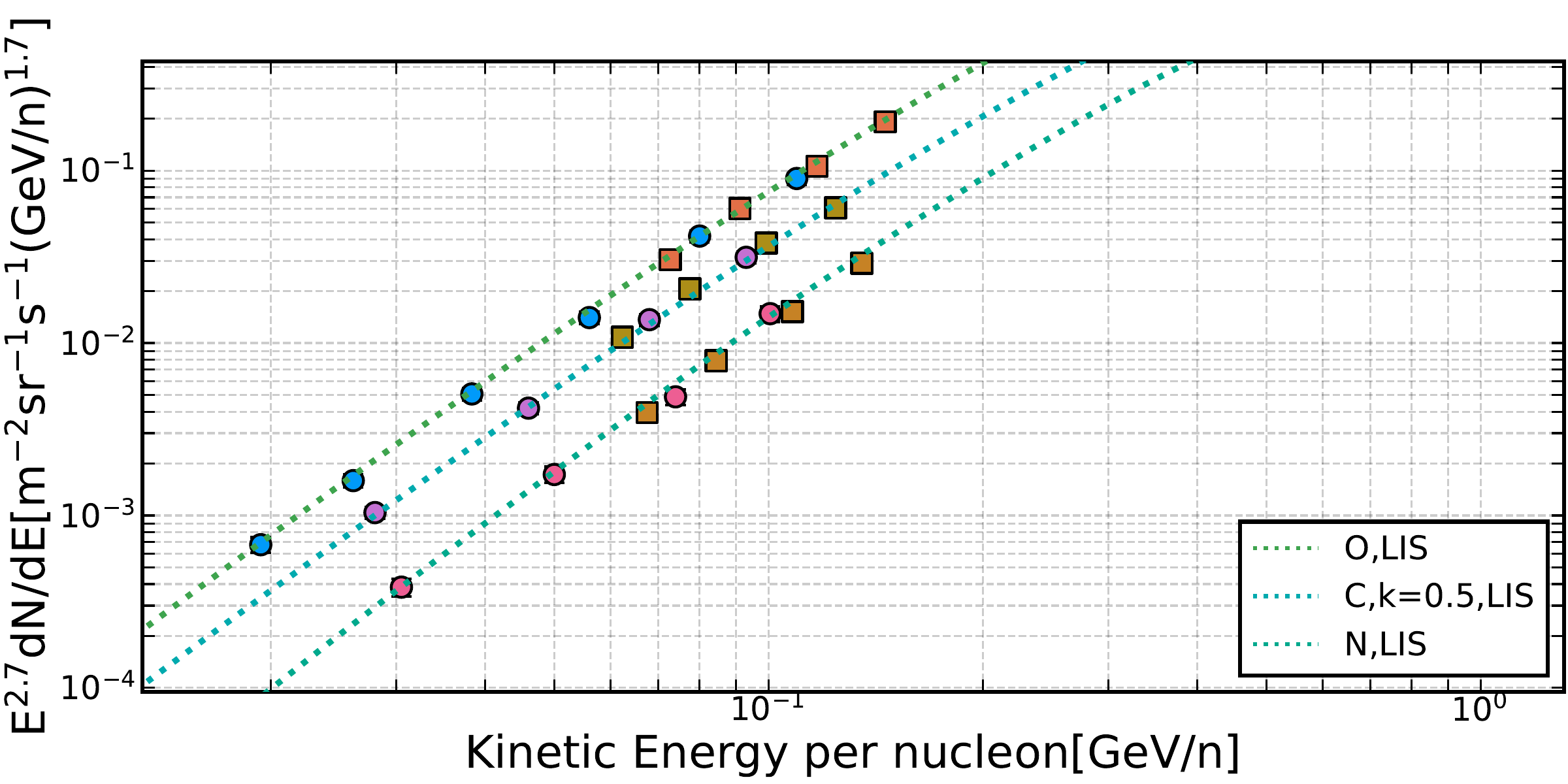}
\includegraphics[width=0.45\textwidth,trim=0 0 0 0,clip]{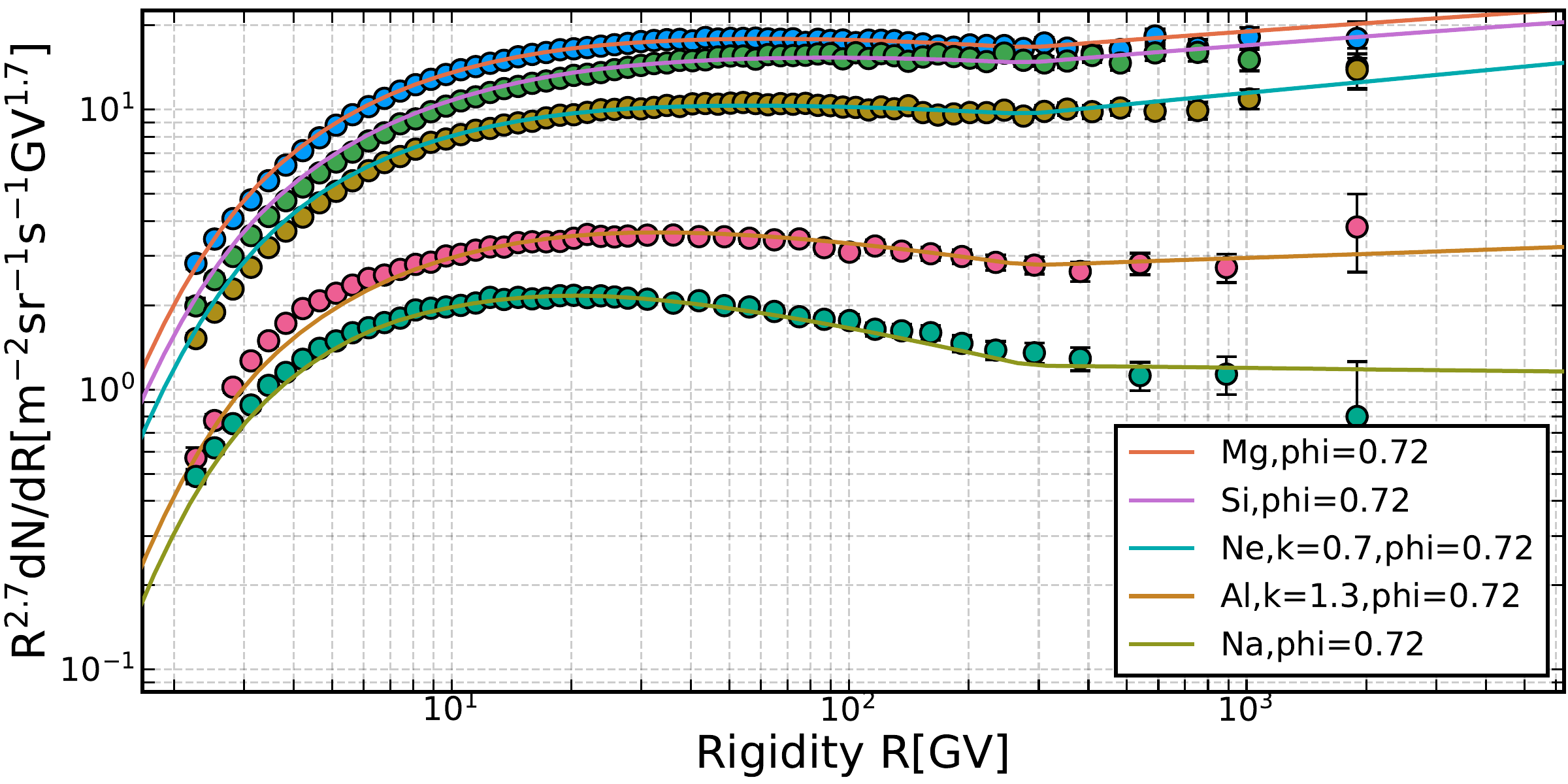}
\includegraphics[width=0.45\textwidth,trim=0 0 0 0,clip]{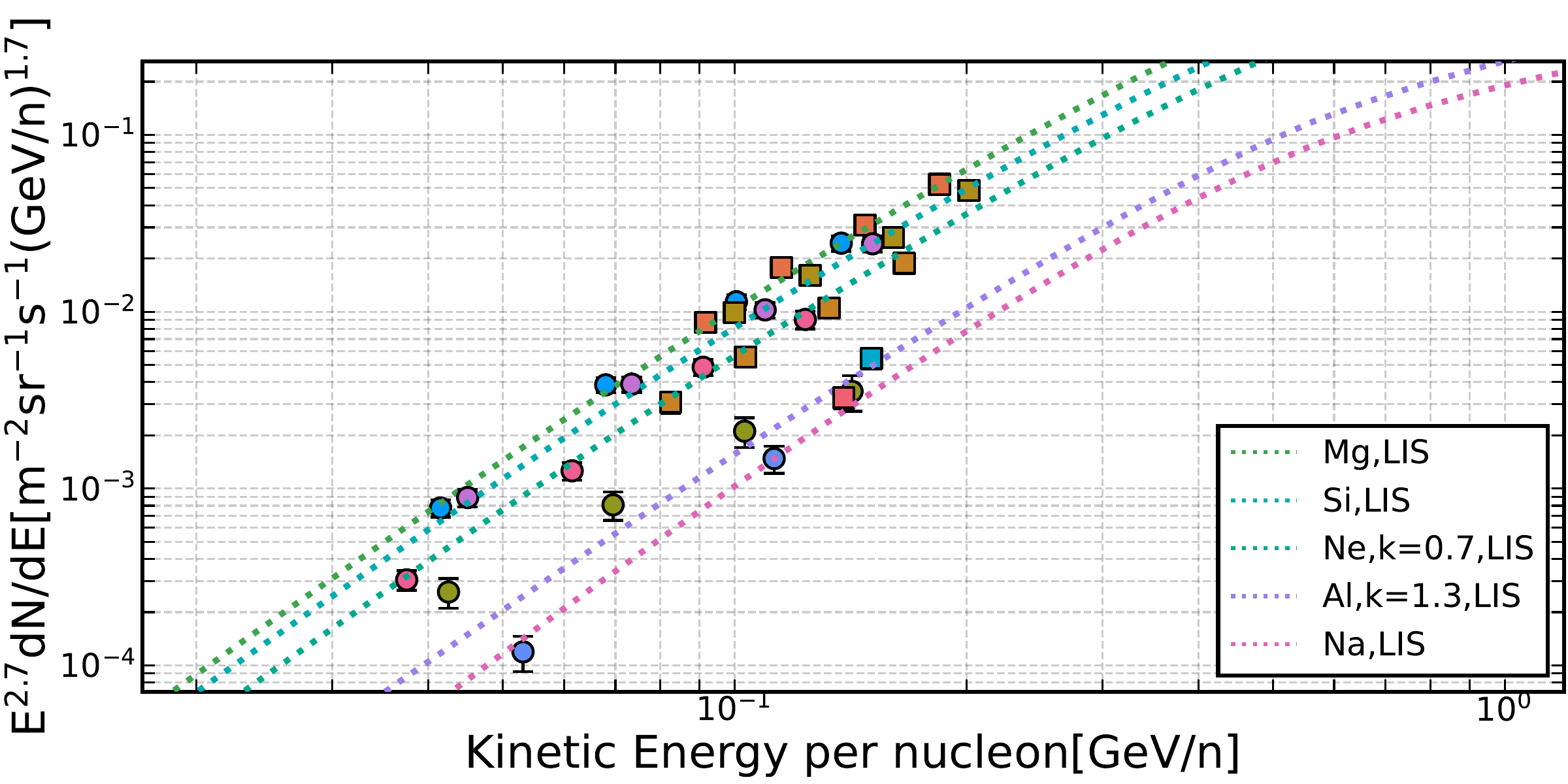}
\includegraphics[width=0.45\textwidth,trim=0 0 0 0,clip]{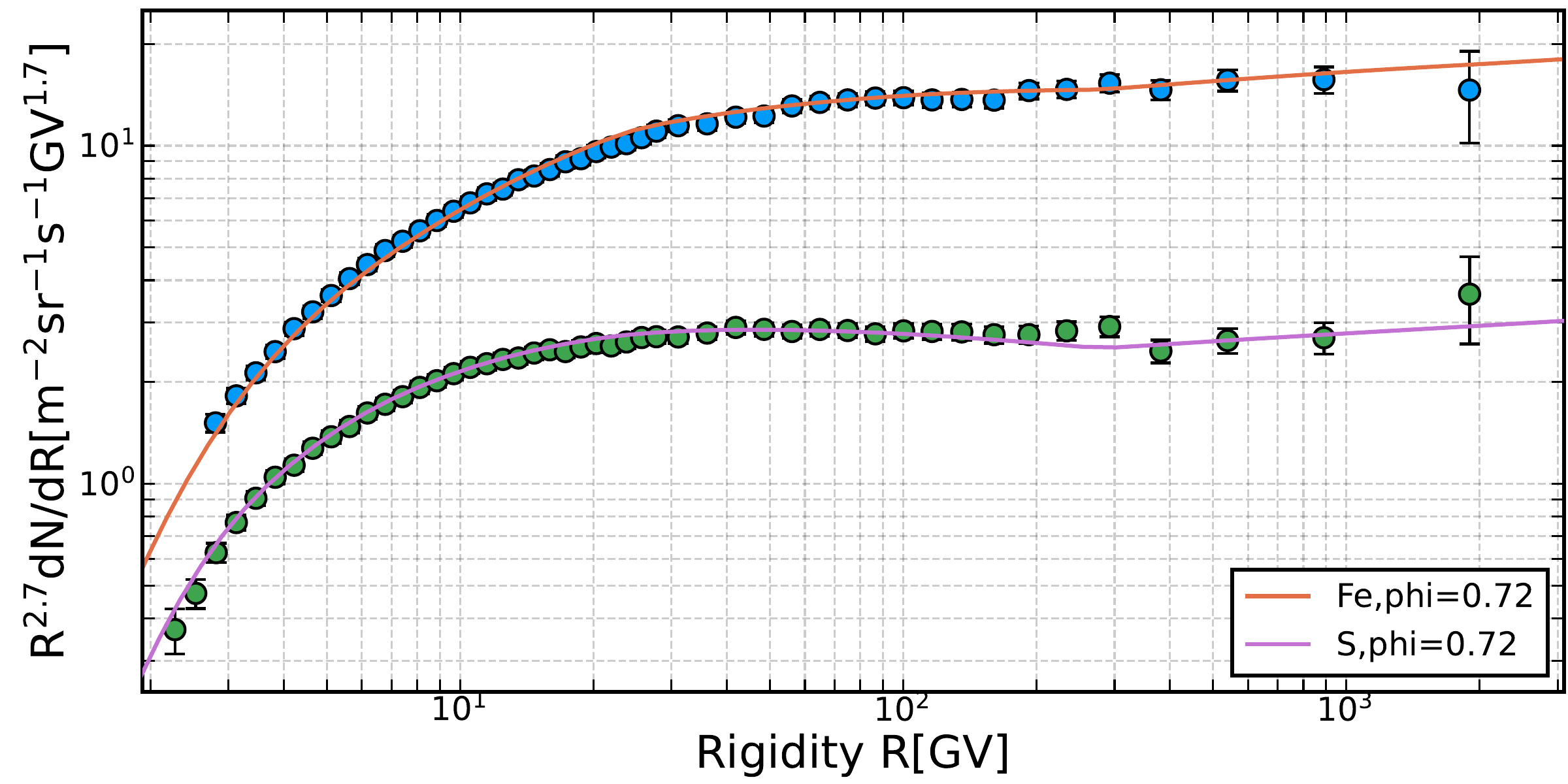}
\includegraphics[width=0.45\textwidth,trim=0 0 0 0,clip]{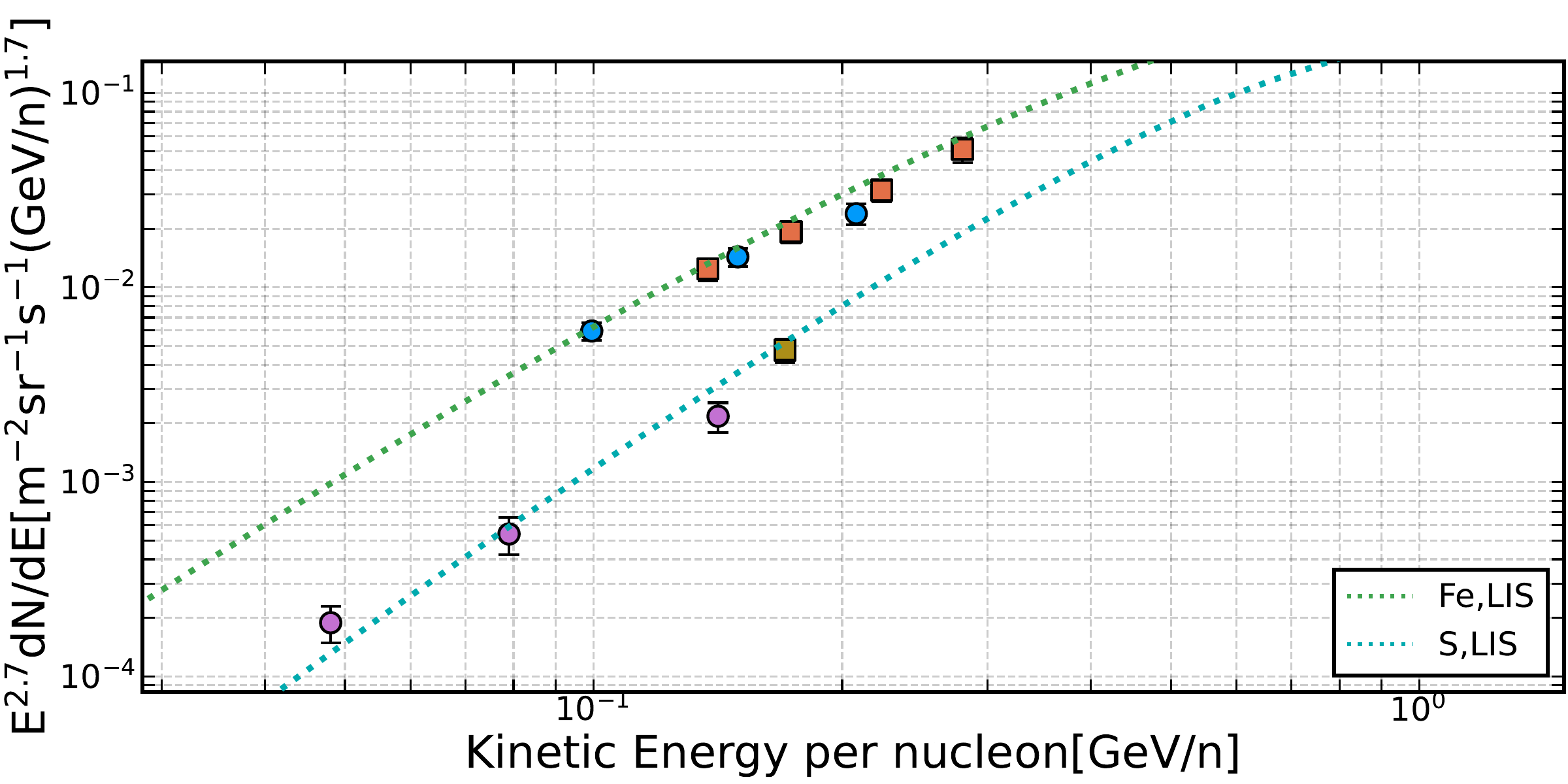}
\caption{\label{fig:primary} 
Left: The primary spectra calculated with the best-fit parameters, compared with the experimental data of AMS-02~\cite{AMS:2020cai, Aguilar:2021tos, AMS:2021brg, AMS:2021lxc, AMS:2023anq}.
Right: The local interstellar spectra calculated with the best-fit parameters, compared with the experimental data of Voyager~\cite{Cummings:2016pdr} (circle: HET-Aend, rectangle: HET-Bend).
Some spectra and data have been rescaled by k for better illustration.
}
\end{figure*}

To describe the common bump structure of nuclei fluxes at low rigidities \cite{Johannesson:2016rlh, Phan:2021iht}, we model the injection spectrum of primary nuclei as a broken power law, with slope indices $\nu_0$ and $\nu_1$ below and above the low-energy break rigidity $R_{br}$: 
\begin{equation}
\begin{aligned}
   q(R)=\begin{cases} (R/R_{br})^{\nu_0}\,,  \quad& R<R_{br} \\
   (R/R_{br})^{\nu_1}\,,  \quad& R\geq R_{br}\end{cases}\,.
\end{aligned}
\end{equation}
As one main difference to the previous work, we assume that the injection parameters ($\nu_0$, $R_{br}$ and $\nu_1$) can be diverse among primary species, to better fit their corresponding spectra \cite{AMS:2020cai, Aguilar:2021tos, AMS:2021brg, AMS:2021lxc, AMS:2023anq}.
The constrained injection parameters and source abundances of primary CRs are listed in Table~\ref{table: injection}.
The best-fit spectral indices $\nu_0$ of mid-mass nuclei C/N/O are equal (1.249), which is softer than that of heavy nuclei from Ne to S (1.1). It implies that the fragmentation loss may be stronger than what {\footnotesize GALPROP} assumed for heavy nuclei, or these CRs may be generated and accelerated at different types of SNR sources. 
The break rigidity $R_{br}$ of Fe is much larger than that of other nuclei, resulting in a significant reduction of the spectrum below 25~GV, possibly due to the underestimation of fragmentation loss during propagation.
The spectral index $\nu_1$ above the break rigidity slightly changes with nuclei mass, but the differences are hard to distinguish due to large data uncertainties.
In Figure~\ref{fig:primary} we illustrate the best-fit primary spectra of different species compared with observations from AMS-02~\cite{AMS:2020cai, Aguilar:2021tos, AMS:2021brg, AMS:2021lxc, AMS:2023anq} (left panels) and Voyager~\cite{Cummings:2016pdr} (right panels).
In the center-left panels, the primary spectra Ne/Na/Mg/Al/Si are constrained by the AMS-02 measurements with small inconsistencies at $\sim5$~GV. The overshoot of Ne and Si spectra suggests a smaller spectral index of their injection spectra or the underestimation of fragmentation loss. The excess of Al and Na spectra was also found in Ref.~\cite{Boschini:2022fpd}, which could be the extra contribution from local Wolf-Rayet stars.

To consider the solar modulation effect on the spectrum inside the heliosphere, we adopt the force-field approximation \cite{Gleeson:1968zza}. The strength is described by the solar modulation potential $\phi$.
Most of the AMS-02 measurements used to constrain the parameters were taken during May 2011-May 2018, and measurements of Fe, Al, and Na were taken during May 2011–Oct 2019, hence we assume that these CRs were modulated with the same solar modulation field $\phi$ for simplicity. 
The spectra at low energies depend mostly on the modulation field $\phi$, the low-energy random-walk factor $\eta$, and the injection index $\nu_0$.
To break the degeneracy of these parameters, the local interstellar spectra (LIS) are needed for determining the unmodulated ($\phi=0$) spectra at low energies. 
We include the LIS measurements of different primary CRs from Voyager 1~\cite{Cummings:2016pdr,webber2018measurementsinterstellarspacegalactic} in the fitting, and the estimated modulation field $\phi$ is 0.72$\pm0.05$~GV.

For the resolution of the {\footnotesize GALPROP} calculation in the work, we set a 2D spatial grid of dr=1~kpc and dz=0.2 kpc, and an energy grid of Ekin\_factor=1.2, considering both accuracy and speed. Other parameters in the paper are kept as the defaults of {\footnotesize GALPROP} v57, see the \texttt{galdef\_57\_CRfit\_Pulsar\_varXCO} example provided in the code.

\subsection{Cross section\label{sec:xs}}

\subsubsection{modification}
The default cross-section models developed in {\footnotesize GALPROP} code are [GAL12] and [GAL22], partly based on their fits of a compilation of cross-section measurements and code evaluations, and partly based on the Webber's \cite{Webber:1990pr} or Silberberg's \cite{Silberberg:1998lxa} parametrization with semi-empirical formulas.
Since the propagation parameters were determined in Ref.~\cite{Zhao:2024qbj} by fitting the B/C and B/O ratio, which was based on the [GAL12] model, we adopted the same model to calculate the production of Li.
For important reactions, the cross sections are mostly decided by the data sheet named \texttt{eval\_iso\_cs.dat} in {\footnotesize GALPROP} code. We have updated this data sheet by including more observations, which are introduced in Appendix~\ref{app:xsdata}.

It is important to note that Webber's parametrization (hereafter [WE93] for short) cannot predict the cross sections of $Z<4$ fragments, such as $\rm^6Li$, $\rm^7Li$ and $\rm^6He$\footnote{The short-lived intermediate nuclei $\rm^6He$ will decay quickly into $\rm^6Li$ before they can collide on the ISM gas.}.
Only dominant contributions from C, N, O, and Mg were calculated in [GAL12], either by following a direct fit to the available data or by adopting Silberberg's parametrization (hereafter [TS00]). The production of $\rm^6Li$ and $\rm^7Li$ from other subdominant projectiles (e.g. Ne, Si, and Fe) are not accounted for, and the production of ghost nucleon $\rm^6He$ from almost all the projectiles are not accounted for either.
Because of that, the well-known lithium excess~\cite{Boschini:2019gow, Weinrich:2020ftb, DeLaTorreLuque:2021yfq} can be significantly biased if implementing the cross-section model based on [WE93].

\begin{figure}[htbp]
\includegraphics[width=0.5\textwidth,trim=0 0 0 0,clip]{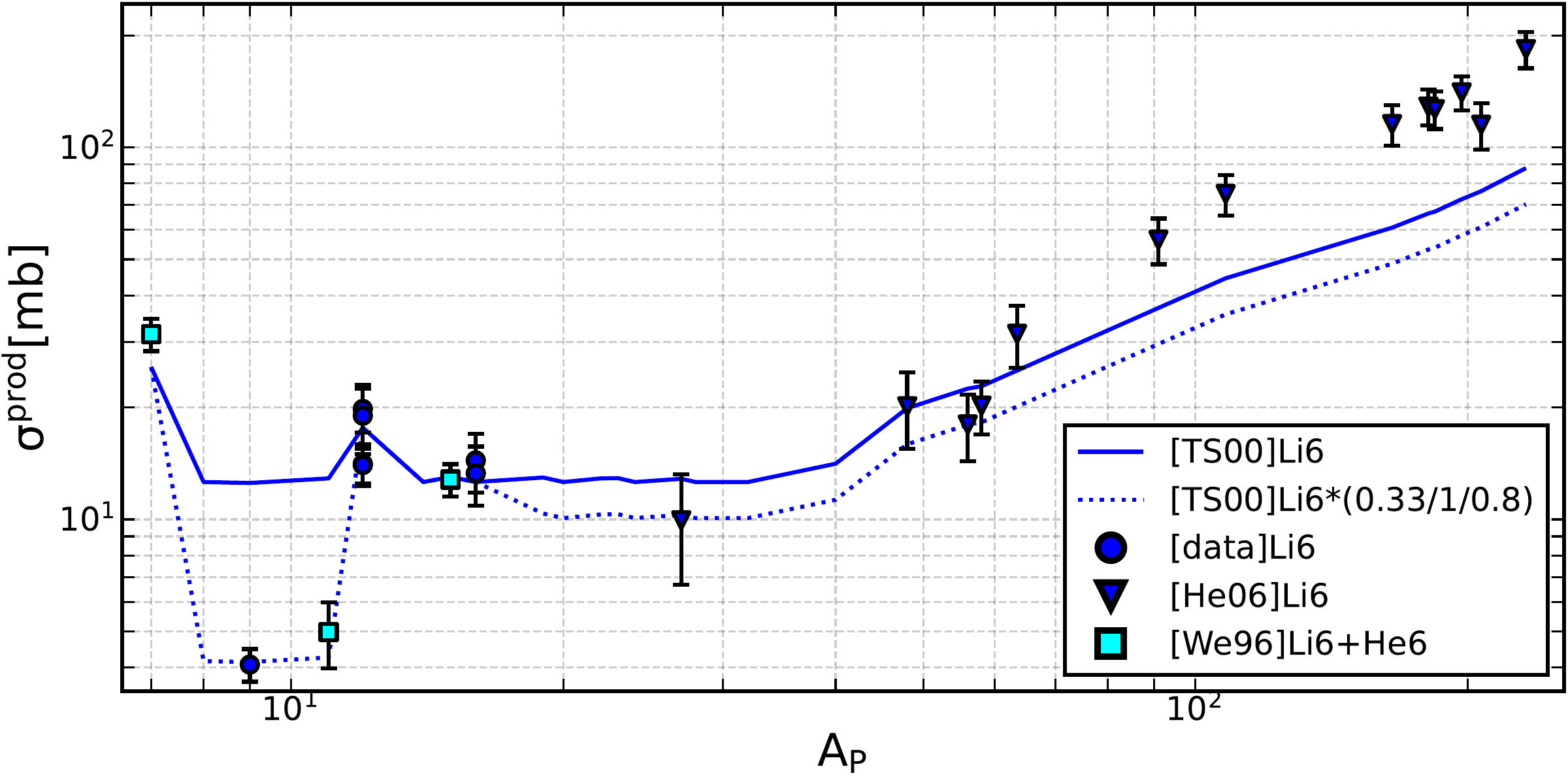}
\includegraphics[width=0.5\textwidth,trim=0 0 0 0,clip]{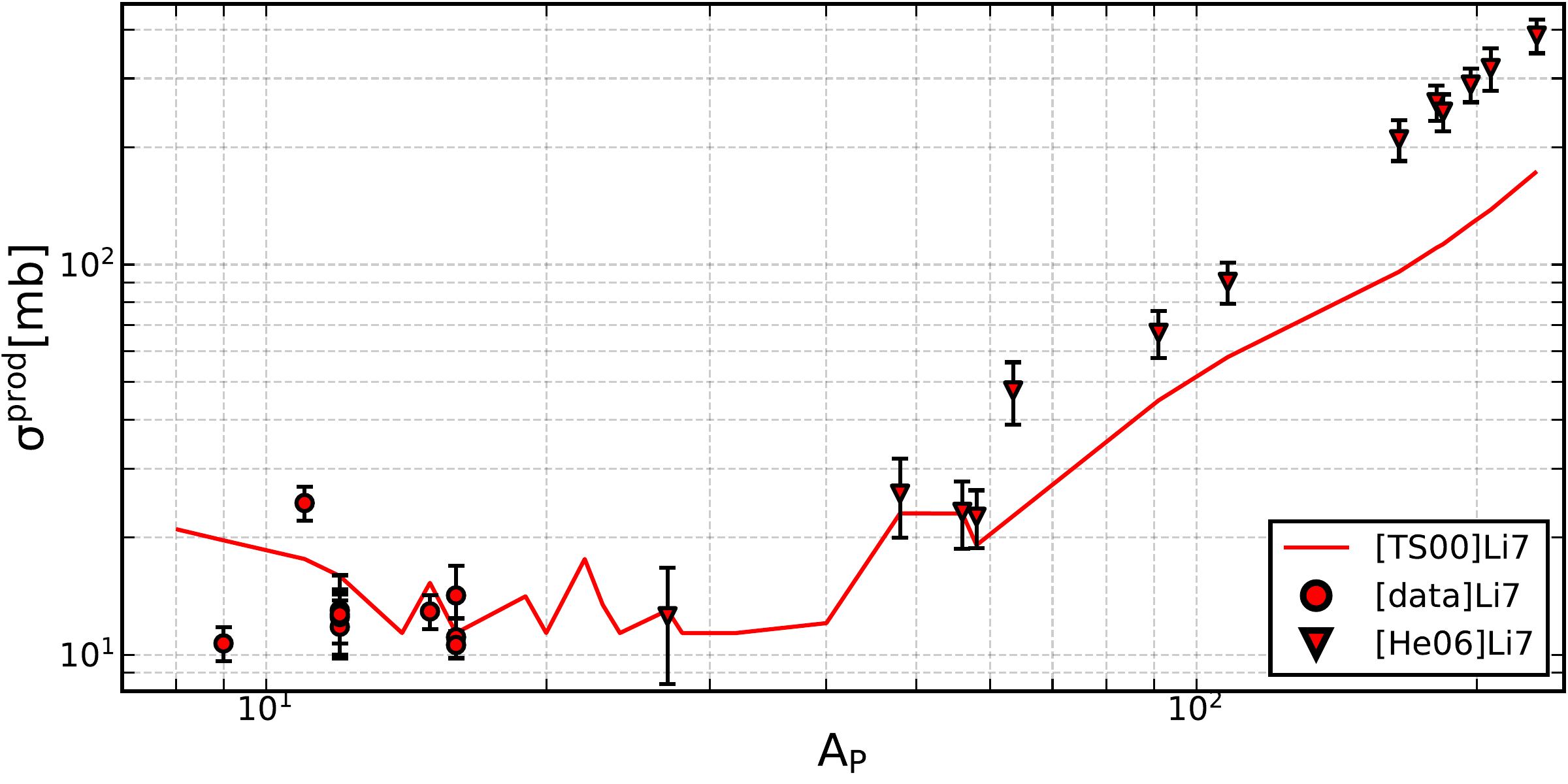}
\includegraphics[width=0.5\textwidth,trim=0 0 0 0,clip]{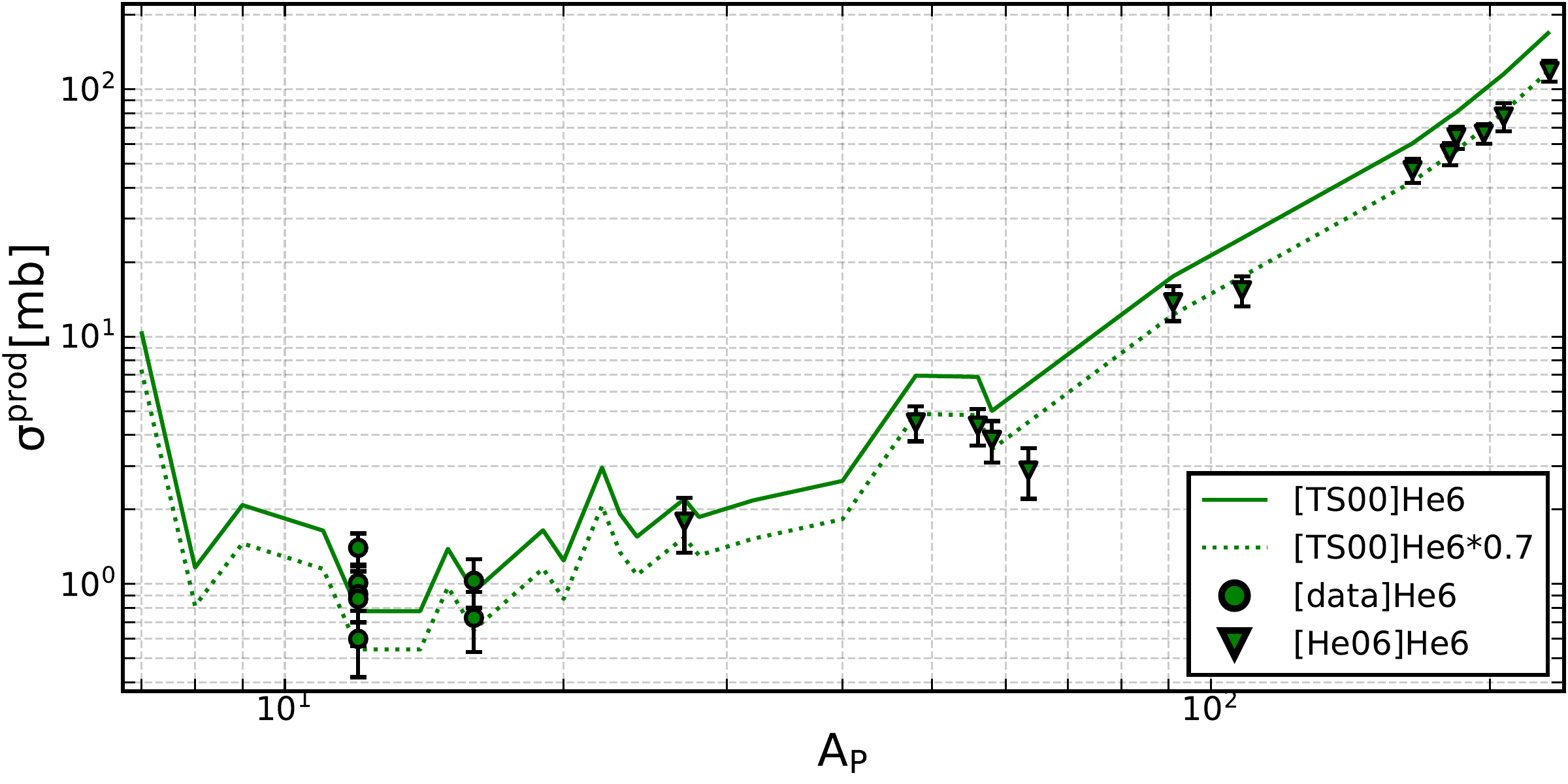}
\caption{\label{fig:xsscale_li} 
The cross sections of $\rm^6Li$ (top), $\rm^7Li$ (center), and $\rm^6He$ (bottom) as a function of the progenitor mass based on the parametrization of [TS00]~\cite{Silberberg:1998lxa}.
The dotted lines represent the modified cross-section results used in the work.
[He06] represents new data from Ref.~\cite{Herbach:2006rw}, [data] represents the data compilation provided by {\footnotesize GALPROP}, and [We96] represents the cumulative cross sections ($\rm^6Li$+$\rm^6He$) at 400~MeV/$n$~\cite{Ramaty:1996qt}.
}
\end{figure}

In this work, we included the missing reactions not considered in [WE93] by using the parametrization of [TS00], as was suggested in Refs.~\cite{Evoli:2017vim, Maurin:2022irz}. After adding missing reactions, the change is obvious for the Li production at all energies, while negligible for other secondaries. As a preliminary result, we noticed a significant overestimation of Li compared with the AMS-02 observations~\cite{Aguilar:2021tos}, which was also found in Ref.~\cite{Maurin:2022irz}.

Most of the added reactions are not measured, meaning that the parametrization results would undershoot or overshoot the real values systematically.
In Figure~\ref{fig:xsscale_li}, we illustrate the cross sections of $\rm^6Li$, $\rm^7Li$, and $\rm^6He$ as a function of the progenitor mass $\rm A_P$ based on the calculation of [TS00] parametrization.
To verify the prediction of [TS00], we draw the available measurements in the figure, for details of the datasets we refer the reader to the Appendix~\ref{app:xsdata}.
Thanks for the new cross-section data of Ref.~\cite{Herbach:2006rw} (triangle points) from the fragmentation of heavy nuclei ($A\geq27$) to the $\rm^6Li$, $\rm^7Li$ and $\rm^6He$, we discovered a global overestimation of the unmeasured cross sections from heavy projectiles.
In the top panel, the cross-section measurements of light projectiles also imply a strong overestimation that should be systematically reduced.
According to the detailed analysis in Appendix~\ref{app:renorm}, we apply these modifications to the cross-section results of [TS00]:
\begin{enumerate}
\item For $\rm^6Li$ production: 
The cross sections of $Z>8$ projectiles are rescaled by $80\%$, except for $\rm^{27}Al$, $\rm^{48}Ti$, $\rm^{56}Fe$ and $\rm^{58}Ni$.

\item For $\rm^6Li$ production: 
The cross sections of $Z<6$ projectiles are rescaled by $33\%$, except for $\rm^{7}Li$, $\rm^{9}Be$ and $\rm^{11}B$.

\item For $\rm^6He$ production: 
The cross sections of all projectiles are rescaled by $70\%$, except for $\rm^{14}N$, $\rm^{12}C$, $\rm^{16}O$, $\rm^{27}Al$, $\rm^{48}Ti$, $\rm^{56}Fe$ and $\rm^{58}Ni$.
\end{enumerate}
The production of $\rm^7Li$ is not modified since the available data fit well with [TS00]'s prediction (center panel).
The cross sections of other reaction channels are automatically renormalized to their available measurements and, hence, are excluded in the modifications mentioned above.

\subsubsection{ranking of contribution\label{sec:rank}}
It is essential to quantify the uncertainties associated with each production channel.
Here we follow a data analysis routine introduced in our previous work \cite{Zhao:2022bon, Zhao:2024qbj}, which was used to analyze the cross-section uncertainty of the F, B, and Be production. 

Firstly, we list the important reaction channels of Li isotopes arranged by their contribution rates in Tables~\ref{tab:channel} and~\ref{tab:channel2}.
The contribution rates are calculated by the fraction ratio of each channel with 
\begin{equation}
	f_{abc}=\frac{\psi-\psi(\sigma^{a+b\rightarrow c}=0)}{\psi}\,,
\end{equation}
where $\psi$ is the flux of isotope Li, and the target b is the ISM gas which includes the $\rm^4He$-target contributions.
Half of the secondary Li fluxes are produced from the main projectiles of $^{12}\text{C}$ and $^{16}\text{O}$, as they have the richest abundances (see Table~\ref{table: injection}).
The $\rm^6He$ contributions from $^{12}\text{C}$ and $^{16}\text{O}$ are relatively smaller but non-negligible ($\sim1.4\%$ and $\sim1.8\%$), as the $\rm^6He$ cross sections are about one magnitude smaller than that of $\rm^6Li$.
The Li contributions of N, Ne, Mg, Si, and Fe are subdominant since they have poorer abundances, and the cross sections of the related channels are less or not constrained. We also list the summarized contributions of heavy projectiles $\sum (Z>8)$, which are as much as that of $^{12}\text{C}$ and $^{16}\text{O}$. These reactions were not accounted for in most of the previous works, which could lead to significant underestimation of both $\rm^6Li$ and $\rm^7Li$.

Besides the primary nuclei as the progenitors of Li, we list the secondary nuclei (e.g. $\rm^7Li$, $\rm^{13}C$, $\rm^{11}B$) that can fragment to isotope Li by multistep reactions.
Secondary CRs are firstly produced during the transportation of primary CRs inside the ISM, and they would collide on the ISM to produce Li.
The spectra of B and Be are softer than that of C and O by the slope index of diffusion coefficient $\delta$, and the two-step spectra of Li will be softer than the one-step spectra by $\delta$. Because of that, the production from secondary nuclei can contribute additional bumps to the $\rm^6Li$ and $\rm^7Li$ spectra, which should be taken more care of if we discover some anomalies at low energies.
The cross sections of the related channels are poorly measured and might be overestimated by the available parametrization, according to the analysis in Appendix~\ref{app:renorm}.
Important reactions, especially the peripheral reactions (the projectiles only lose a proton or a neutron) like $^{7}\text{Li}+p \longrightarrow ^{6}\text{Li}$ and $^{7}\text{Li}+p \longrightarrow ^{6}\text{He}$ should be constrained precisely to ensure the reliability of parametrization.

\begin{table}
\caption{\label{tab:channel}Ranking of the $\rm^6Li$ isotopic production cross sections at 10 GeV$/n$, those with contribution rates less than 1\% are not shown. Meanings of cross-section qualities: constrained (A), low-energy constrained (B), renormalized (C), no exp. (D).}
\begin{ruledtabular}
\begin{tabular}{ccc}
 Channel& contribution [\%]&quality\\ \hline
   $^{16}\text{O} \longrightarrow ^{6}\text{Li}$&27.478 &A(12.7\%)\\
  $^{12}\text{C} \longrightarrow ^{6}\text{Li}$&26.022 &A(8.3\%)\\
  $^{56}\text{Fe} \longrightarrow ^{6}\text{Li}$&4.608 &C(20.8\%)\\
  $^{14}\text{N} \longrightarrow ^{6}\text{Li}$&3.241 &B(25\%)\\
  $^{28}\text{Si} \longrightarrow ^{6}\text{Li}$&3.163 &D\\
  $^{24}\text{Mg} \longrightarrow ^{6}\text{Li}$&3.040 &D\\
  $^{20}\text{Ne} \longrightarrow ^{6}\text{Li}$&2.201 &D\\
  $^{16}\text{O} \longrightarrow  ^{6}\text{He}$&1.792 &C(17.8\%)\\
  $\rm^{12}C \longrightarrow ^{6}He$&1.403 &A(10.8\%)\\
\hline
  $^{7}\text{Li} \longrightarrow ^{6}\text{Li}(^{6}\text{He})$\footnote{The cross sections of ghost nucleon $^{6}\text{He}$ produced from $^{7}\text{Li}$, $^{11}\text{B}$ and $\rm{^{15}N}$ are added to that of $^{6}\text{Li}$, since the data \cite{Ramaty:1996qt} used to renormalize them are cumulative cross sections. See also in App.~\ref{app:xsdata}.}&4.148 &C(20\%)\\
  $^{13}\text{C} \longrightarrow ^{6}\text{Li}$&2.997 &D\\
  $^{15}\text{N} \longrightarrow ^{6}\text{Li}(^{6}\text{He})$&2.614 &C(20\%)\\
  $^{11}\text{B} \longrightarrow ^{6}\text{Li}(^{6}\text{He})$&1.410 &C(26.4\%)\\
\hline
$\rm\sum (Z>8) \longrightarrow ^{6}Li$&21.431 &D\\
$\rm\sum (Z>8) \longrightarrow ^{6}He$&3.302 &D\\
\end{tabular}
\end{ruledtabular}
\end{table}

\begin{table}
\caption{\label{tab:channel2} Same as Table~\ref{tab:channel} but for $\rm^7Li$.}
\begin{ruledtabular}
\begin{tabular}{ccc}
 Channel& contribution [\%]&quality\\ \hline
  $^{12}\text{C} \longrightarrow ^{7}\text{Li}$&26.232 &A(6.6\%)\\
  $^{16}\text{O} \longrightarrow ^{7}\text{Li}$&24.577&A(6.3\%)\\
  $^{56}\text{Fe} \longrightarrow ^{7}\text{Li}$&6.703 &C(20.2\%)\\
  $^{28}\text{Si} \longrightarrow ^{7}\text{Li}$&4.008 &D\\
  $^{28}\text{Mg} \longrightarrow ^{7}\text{Li}$&3.851 &D\\
  $^{20}\text{Ne} \longrightarrow ^{7}\text{Li}$&2.788 &D\\
  $^{14}\text{N} \longrightarrow ^{7}\text{Li}$&2.692 &B(44\%)\\
  $^{22}\text{Ne} \longrightarrow ^{7}\text{Li}$&1.761 &D\\
\hline
  $^{11}\text{B} \longrightarrow ^{7}\text{Li}$&5.657 &C(20\%)\\
  $^{13}\text{C} \longrightarrow ^{7}\text{Li}$&4.170 &D\\
  $^{15}\text{N} \longrightarrow ^{7}\text{Li}$&2.958 &C(20\%)\\
  $^{10}\text{B} \longrightarrow ^{7}\text{Li}$&1.706 &D\\
\hline
  $\rm\sum (Z>8) \longrightarrow ^{7}Li$&30.302 &D\\
\end{tabular}
\end{ruledtabular}
\end{table}

In Tables~\ref{tab:channel} and~\ref{tab:channel2}, we score the cross-section qualities of these channels based on the available measurements, which can be used to estimate the cross-section uncertainties and as a reference for facilities that can perform relevant measurements.
Quality A means that many experiments can be used to determine the cross sections at all energies by interpolation.
Quality B means that the experiments are enough to determine the low-energy cross sections, but extrapolation should be made to predict the cross sections at multi-GeV energies.
Quality C means the parametrization result (from [WE93] or [TS00]) is renormalized to only one or two data.
Quality D means the parametrization result is taken at face value without constraints from available experiments.

The errors of A-quality and B-quality channels are derived by making least-squares fits to the cross-section data and determining the Gaussian-distributed dispersion.
As introduced in our previous work~\cite{Zhao:2024qbj}, the prediction of secondary nuclei relies seriously on the extrapolation of the high-energy data measured by multiple experiments. For A-quality channels, we only choose the data above 2 GeV/$n$ to determine the uncertainties in the extrapolation region.

The errors of D-quality channels can be directly taken from Table~9A of Ref.~\cite{Silberberg:1973jxa}.
For the production of Li isotopes, the standard deviation of the [TS00] calculation is (${+30\%}$, ${-25\%}$).
The cross sections of Li from different progenitors are globally calculated by specified formulae constructed in the code. Without constraints from measurements, these results should be systematically correlated with each other.

To estimate the errors of C-quality channels, we construct a simple formula considering both data and parametrization:
\begin{eqnarray}\label{eq:sigma}
	\left(\frac{\Delta\sigma}{\sigma}\right)_C=\sqrt{\delta_{exp}^2+(\frac{0.25}{1+(E_{kin/n}/E_{th})^2})^2}\,,
\end{eqnarray}
where $\delta_{exp}$ and $E_{kin/n}$ are the error and kinetic energy per nucleon of the measurement used to renormalize the parametrization, 0.25 is taken from the standard deviation of the [TS00] calculation. The threshold energy $E_{th}$ is set to $0.6~$GeV/$n$ to ensure that, the error can approach the asymptotic value of $\delta_{exp}$ if the measurement is close to the energy-independent plateau.

The error propagation formula can be used to obtain the total cross-section uncertainties of the secondary nuclei. Following the correlated/uncorrelated uncertainty case in Ref.~\cite{Genolini:2018ekk}, we construct a new formula that is suitable for adding up the cross-section errors of different quality channels:
\begin{eqnarray}\label{eq:mixerror}
	\frac{\Delta\psi_{tot}}{\psi_{tot}}\approx f_{sec}\sqrt{\sum_{A,B,C}\left(f_{abc}\frac{\Delta\sigma}{\sigma}\right)^2+\left(\sum_{D}f_{abc}\frac{\Delta\sigma}{\sigma}\right)^2}\,,
\end{eqnarray}
where $f_{sec}$ is the secondary fraction of the nuclei which is 100\% for pure secondary particles, $f_{abc}$ and $\frac{\Delta\sigma}{\sigma}$ are the contribution ratio and cross-section error of each reaction, which are given in Tables~\ref{tab:channel} and~\ref{tab:channel2}.
The error of D-quality channels is set to (${+30\%}$, ${-25\%}$) for the production of Li isotopes, and they are added linearly.
In Eq.~\ref{eq:mixerror}, the total uncertainty is described by the root squared sum of an uncorrelated component and a correlated component.
We assumed that the uncertainties of A, B, and C-quality reactions are uncorrelated since they are constrained by different measurements, while the uncertainties of D-quality reactions are correlated.

The data uncertainties of Li isotopes is about 5\% for the forthcoming AMS-02 measurements, while the calculated cross-section uncertainty of $\rm^6Li$ is (+9.2\%, -8.1\%), and that of $\rm^7Li$ is (+9.9\%, -8.4\%).
We notice that D-quality channels in Tables~\ref{tab:channel} and~\ref{tab:channel2} are unmeasured and correlated, hence the cross-section determination of these channels by new experiments should be helpful.
If the contributions of subdominant projectiles, such as Ne/Mg/Si, are measured at high energies (for example $\pm10\%$ at 2~GeV/$n$), the related channels would be uncorrelated and changed to C-quality. Therefore, the cross-section uncertainties of $\rm^6Li$ can be reduced to (+7\%, -6.3\%) and the uncertainties of $\rm^7Li$ can be reduced to (+6.5\%, -5.7\%), which can be comparable to the CR data uncertainties.
On the other hand, for dominant reactions like $^{12}\text{C}+p \longrightarrow ^{6}\text{Li}$ and $^{16}\text{O}+p \longrightarrow ^{6}\text{Li}$, the errors are still large and can be improved by including more precise measurements.

\section{PREDICTION\label{sec:result}}
The preliminary AMS-02 results on Li isotopes were reported in the International Cosmic Ray Conference~\cite{AMS_icrc2021}, which extend the energy range beyond that of previous experiments with unprecedented precision.
In this section, we calculate the spectra of $\rm^6Li$ and $\rm^7Li$ that can be compared to the preliminary data.
The estimated modulation field $\phi$ is set to 0.72$\pm0.05$~GV, since the preliminary isotope spectra were taken during the same period (May 2011–Oct 2019) as that of other nuclei spectra used in the work.

\begin{figure}[htbp]
\includegraphics[width=0.5\textwidth,trim=0 0 0 0,clip]{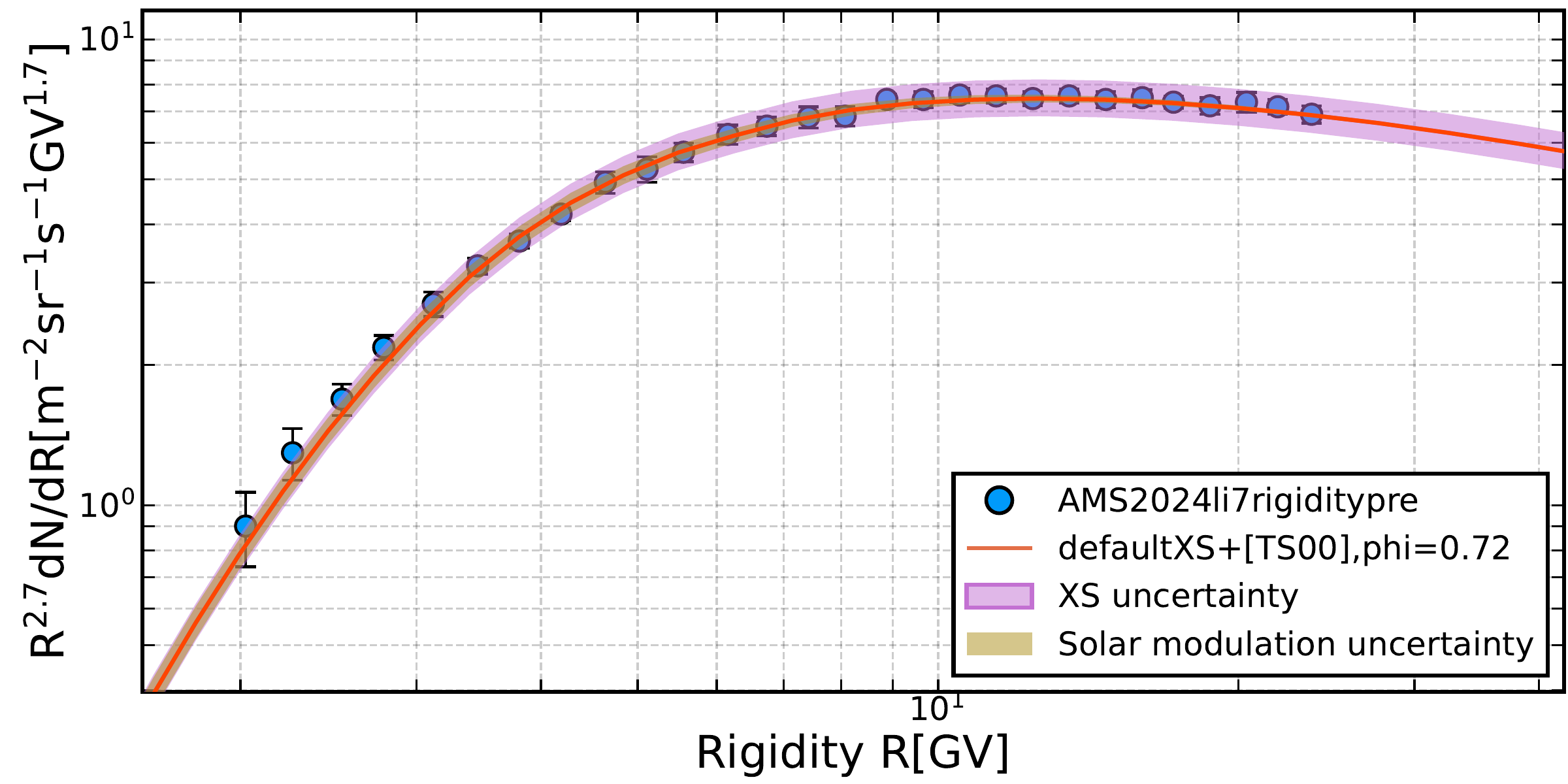}
\includegraphics[width=0.5\textwidth,trim=0 0 0 0,clip]{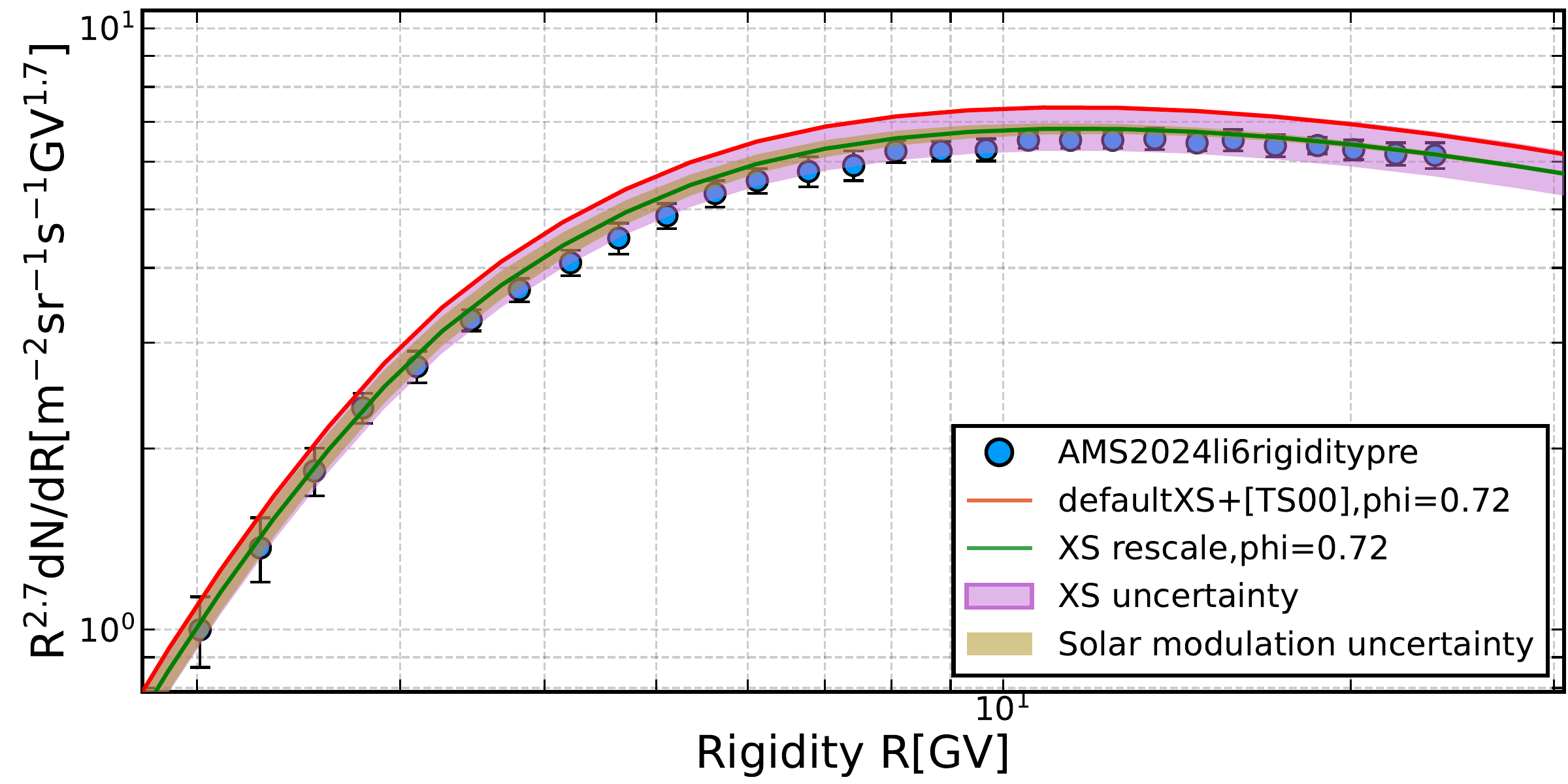}
\caption{\label{fig:isoLI_flux} 
The predicted $\rm^7Li$ and $\rm^6Li$ spectra compared with the preliminary AMS-02 experimental data~\cite{AMS_icrc2021}.
The red lines represent the spectra based on the default cross-section model with extra reactions calculated by [TS00]~\cite{Silberberg:1998lxa}. The green line represents the spectrum with modifications of unmeasured channels.
The purple-shaded band represents the cross-section uncertainties and the brown-shaded band represents the uncertainties of solar modulation potential.
}
\end{figure}

In Figure~\ref{fig:isoLI_flux}, we illustrate our prediction of $\rm^6Li$ and $\rm^7Li$ spectra with two sets of cross-section models.
The first set (red) is based on the [GAL12] model in {\footnotesize GALPROP} with additional reactions calculated by [TS00] parametrization, and the second set (green) includes the modifications applied to the [TS00] results obtained in the previous section.
It is clear that, if adding these missing channels according to [TS00] parametrization, the predicted $\rm^7Li$ spectrum fits well with the AMS-02 result, but the predicted $\rm^6Li$ spectrum is significantly larger over the energy range of the data.
After rescaling the production cross sections of [TS00] results, the calculated $\rm^6Li$ flux (green line) is reduced by at most 8.5\%, which fits better with the AMS-02 result.
It means that these modifications are important for improving the predictions of Li isotopes.
We find that at around 7~GV, the predicted flux is slightly larger by 5\% compared with the preliminary observations.
The slight overestimation can be due to the correlations in the data systematics since the preliminary measurements of AMS-02~\cite{AMS_icrc2021} are given by three radiators at different energy regions. For the NaF radiator with energy from 0.82-4.65~GeV/$n$, the measured $\rm^6Li$ data are systematically smaller compared with that of the TOF radiator ($<$1.11~GeV/$n$) and AGL radiator ($>$3.79~GeV/$n$).

We have also drawn the cross-section and solar modulation uncertainties in Figure~\ref{fig:isoLI_flux}.
The cross-section uncertainties (purple band) are estimated in Section~\ref{sec:rank} to be (+9.2\%, -8.1\%) for $\rm^6Li$ and (+9.9\%, -8.4\%) for $\rm^7Li$, which is derived at 10 GeV/$n$ and constantly extrapolated to lower energies.
The slight overestimation of $\rm^6Li$ at 7~GV is within the cross-section uncertainties, meaning that the misunderstanding of cross sections might lead to this inconsistency.
Compared with the data uncertainties on isotopic fluxes (5\%) above 10~GV, the cross-section uncertainties at present are much larger, hampering us from obtaining meaningful information and distinguishing the source of the inconsistency.
By performing more cross-section measurements on heavy projectiles, the decreased cross-section uncertainties can finally be comparable to the data uncertainties.
The solar modulation uncertainty (brown band) is estimated to be $\pm0.05$~GV, which is only as important as cross-section uncertainties below 2~GV and, thus, should be a subdominant effect.

\begin{figure}[htbp]
\includegraphics[width=0.5\textwidth,trim=0 0 0 0,clip]{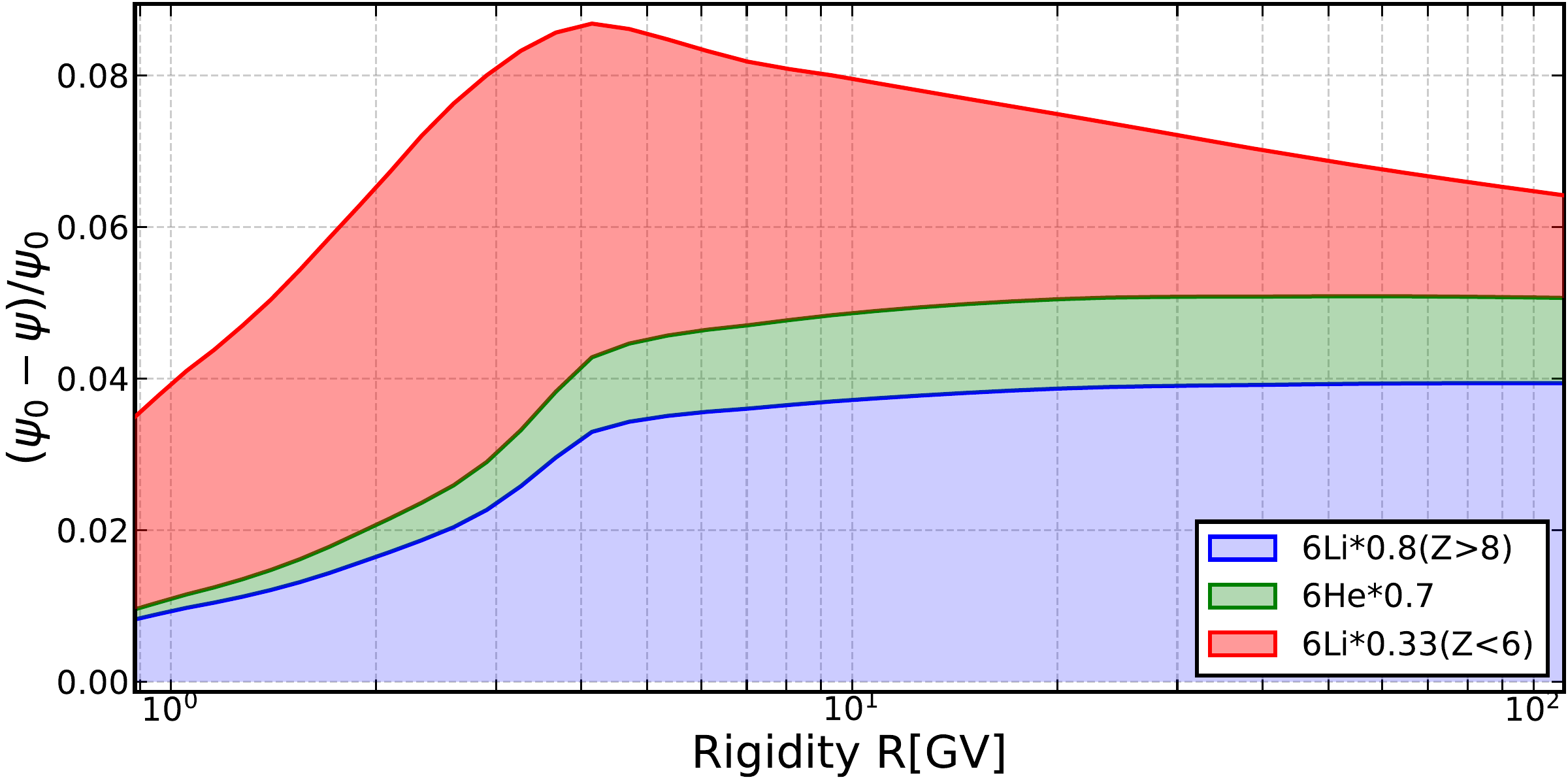}
\caption{\label{fig:li6change} 
The decreasing fraction of the $\rm^6Li$ spectrum caused by three kinds of modifications. Blue band: modification of $\rm^6Li$ production from heavy projectiles. Green band: modification of $\rm^6He$ production.
Red band: modification of $\rm^6Li$ production from light projectiles.
}
\end{figure}
In Figure~\ref{fig:li6change}, we estimate the decreasing fractions $(\psi_0-\psi)/\psi_0$ of the $\rm^6Li$ flux caused by the three kinds of modifications. 
The modification of $\rm^6Li$ production from heavy projectiles (blue band) increases to a constant reduction of 4\% at above 10~GV, which is similar to the features of the related cross sections. The reason is that the dominant contributions of this modification are from primary nuclei such as Ne/Mg/Si with similar spectral indices as that of C/O.
The modification of $\rm^6He$ (green band) has the same feature, but its decreasing fraction is smaller due to generally smaller cross sections of $\rm^6He$.
The modification from light projectiles (red band) caused a huge part of the reduction, without which the predicted $\rm^6Li$ flux would be too large to be favored by the AMS-02 result below 10~GV.
Considering that the spectra produced from light projectiles are softer than those of primary projectiles, the contributions from light projectiles would not impact the high-energy spectrum.
The overestimation of $\rm^6Li$ at 7~GV could be revolved by reducing the production cross sections of $\rm^6Li$ from secondary nuclei. However, most of the production has already been reduced (66\%) in our modification, and we cannot believe that they are still overestimated.

\begin{figure}[htbp]
\includegraphics[width=0.5\textwidth,trim=0 0 0 0,clip]{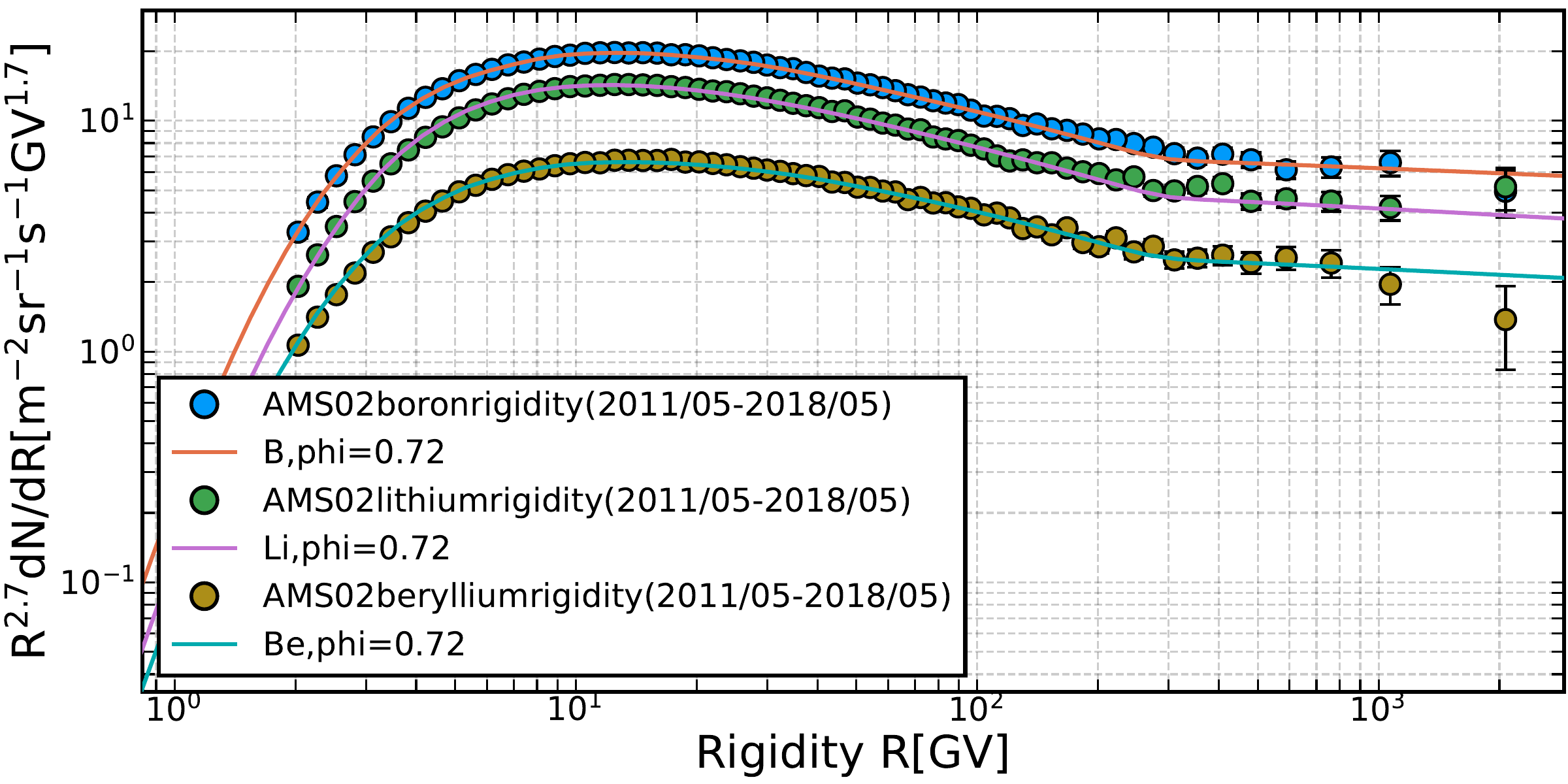}
\includegraphics[width=0.5\textwidth,trim=0 0 0 0,clip]{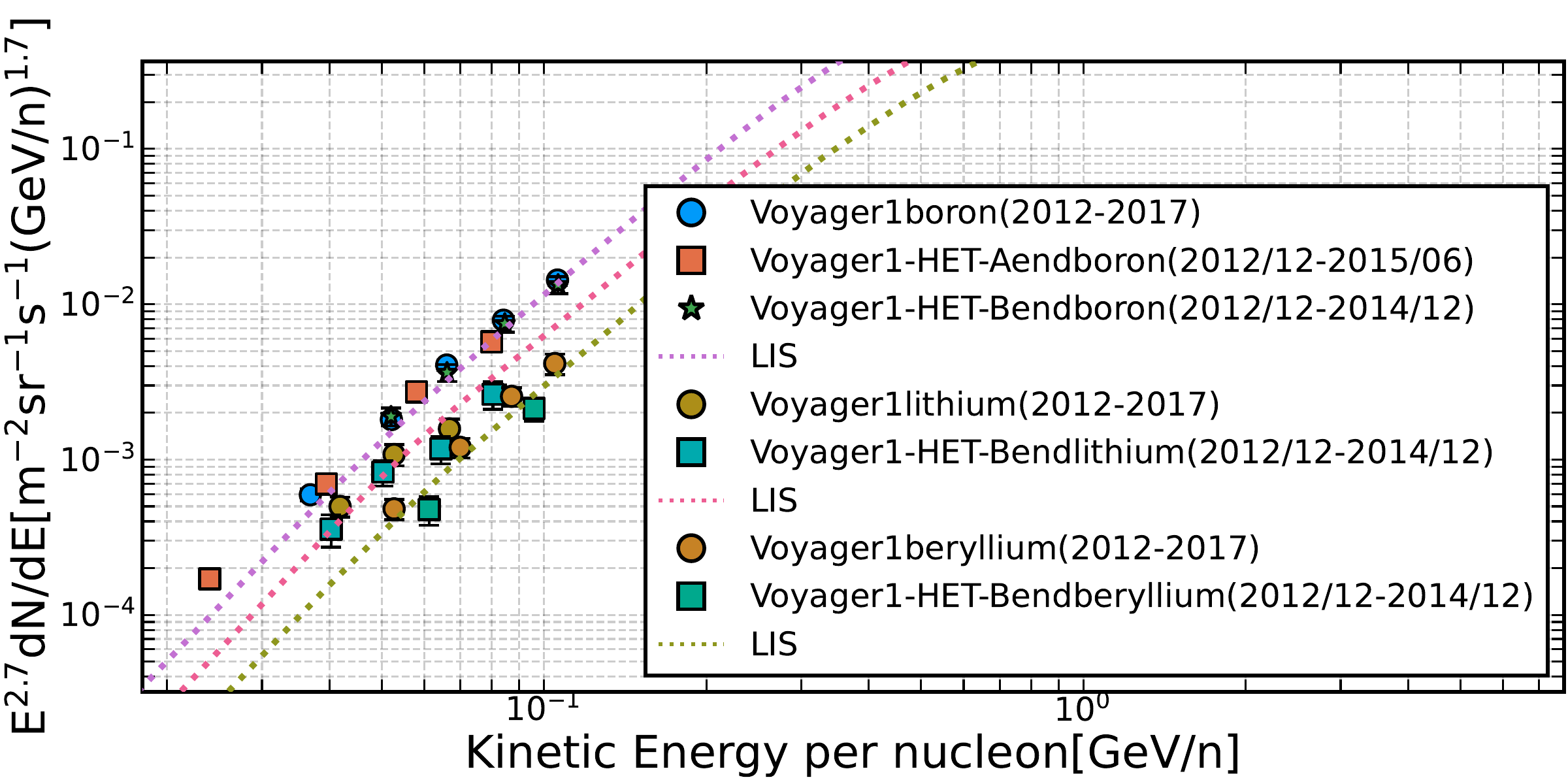}
\caption{\label{fig:LIBEB_flux} 
The secondary spectra calculated with the best-fit parameters, compared with the experimental data: AMS-02~\cite{Aguilar:2021tos} and Voyager~\cite{Cummings:2016pdr,webber2018measurementsinterstellarspacegalactic}. Secondaries: B, Li, and Be spectra from top to bottom.
}
\end{figure}

To check the prediction of secondary particles at higher energies, we illustrate the total fluxes of Li, Be and B to be compared with AMS-02 measurements~\cite{Aguilar:2021tos} in Figure~\ref{fig:LIBEB_flux}. The predicted spectrum of Li is well-constructed by our updated cross-section models, consistent with the total spectra of B and Be. Therefore the propagation consistency is conserved among these CRs.
The LIS of these secondaries are shown in the bottom panel and compared with Voyager's results~\cite{Cummings:2016pdr,webber2018measurementsinterstellarspacegalactic}. These spectra generally pass through the measurements without significant anomalies, and the small deviation can be attributed to the misunderstanding of cross sections below 100~MeV/$n$.

\section{SUMMARY\label{sec:conclusion}}

This paper follows up on our previous work~\cite{Zhao:2024qbj} to solve the overestimation of Li found in Ref.~\cite{Maurin:2022irz} with data-driven cross sections and make predictions to the Li isotope spectra.

We have made several improvements to the default cross-section model of [GAL12] in {\footnotesize GALPROP} by including more measurements.
First of all, we add a list of missing reactions that were not calculated in the default cross-section model by using the semi-empirical parametrization from Tsao and Silberberg.
The cross sections of their parametrization were mostly determined by interpolation or extrapolation of other available channels which were barely constrained by experiments. By analyzing the available measurements, we assume that the production cross sections of $\rm^6Li$ and $\rm^6He$ are commonly overestimated and should be reduced. The cross sections of $Z>8$ projectiles to $\rm^6Li$ are rescaled by 80\%, $Z<6$ projectiles to $\rm^6Li$ are rescaled by 33\%, and all projectiles to $\rm^6He$ are rescaled by 70\%, except for the reactions that have been constrained by experiments.

The preliminary Li isotope spectra provided by AMS-02 have reached an unprecedented energy of 12~GeV/$n$ with 5\% uncertainties, which is important for examining the source of the Lithium anomaly.
Using the CR propagation parameters constrained by the B/C, B/O ratios and Be isotopes, we calculate the predictions of Li isotope spectra.
As a result, the $\rm^7Li$ spectrum fits well with the preliminary result of AMS-02, and the fit of the $\rm^6Li$ spectrum is improved by applying the cross-section modifications.
We found a slight overestimation of the $\rm^6Li$ flux at 7~GV, which can be due to large cross-section uncertainties ($8\%\sim9\%$) waiting to be improved by future experiments. The overestimation also suggests that the correlations in the data systematics should be noticed. The total spectrum of Li, together with that of Be and B, are well-reproduced compared with AMS-02 observations, confirming the propagation consistency of mid-mass secondary CRs.
As an extension, the propagation consistency of Li isotopes is also confirmed according to the $\rm^6Li$ and $\rm^7Li$ results in the work.
Extending the propagation network to the heavier nuclei groups shall be meaningful to solve the F/Si anomaly~\cite{AMS:2020cai, Bueno:2022bdc, Zhao:2022bon} found by AMS-02 and give reliable predictions of P/S and sub-Fe/Fe ratios.

We emphasize the importance of properly considering the production cross sections from projectiles heavier than Oxygen, during the analysis of the secondaries such as Li/Be/B. Projectiles lighter than Carbon are also non-negligible when calculating the production below 10 GeV/$n$.
To make reliable predictions for CR transportation, the cross sections of these reactions should be measured at higher energies.
In this work, the large cross-section uncertainties ($8\%\sim9\%$) of $\rm^6Li$ and $\rm^7Li$ mainly come from the lack of measurements on heavy projectiles. We have shown that, by measuring the cross sections of Li isotopes from Ne/Mg/Si, the cross-section uncertainties can be reduced to around 6\%, which is comparable to the data error of AMS-02.
The production of C and O projectiles to $\rm^6Li$ should also be improved for better constraints.

\acknowledgments
This work is supported by the National Key R\&D Program Grants of China under Grandt No. 2022YFA1604802, 
the National Natural Science Foundation of China under Grants No. 12175248, No. 12342502, No. 12105292, and No. 12393853.  

\bibliography{apssamp}

\begin{thebibliography}{69}%
\makeatletter
\providecommand \@ifxundefined [1]{%
 \@ifx{#1\undefined}
}%
\providecommand \@ifnum [1]{%
 \ifnum #1\expandafter \@firstoftwo
 \else \expandafter \@secondoftwo
 \fi
}%
\providecommand \@ifx [1]{%
 \ifx #1\expandafter \@firstoftwo
 \else \expandafter \@secondoftwo
 \fi
}%
\providecommand \natexlab [1]{#1}%
\providecommand \enquote  [1]{``#1''}%
\providecommand \bibnamefont  [1]{#1}%
\providecommand \bibfnamefont [1]{#1}%
\providecommand \citenamefont [1]{#1}%
\providecommand \href@noop [0]{\@secondoftwo}%
\providecommand \href [0]{\begingroup \@sanitize@url \@href}%
\providecommand \@href[1]{\@@startlink{#1}\@@href}%
\providecommand \@@href[1]{\endgroup#1\@@endlink}%
\providecommand \@sanitize@url [0]{\catcode `\\12\catcode `\$12\catcode
  `\&12\catcode `\#12\catcode `\^12\catcode `\_12\catcode `\%12\relax}%
\providecommand \@@startlink[1]{}%
\providecommand \@@endlink[0]{}%
\providecommand \url  [0]{\begingroup\@sanitize@url \@url }%
\providecommand \@url [1]{\endgroup\@href {#1}{\urlprefix }}%
\providecommand \urlprefix  [0]{URL }%
\providecommand \Eprint [0]{\href }%
\providecommand \doibase [0]{https://doi.org/}%
\providecommand \selectlanguage [0]{\@gobble}%
\providecommand \bibinfo  [0]{\@secondoftwo}%
\providecommand \bibfield  [0]{\@secondoftwo}%
\providecommand \translation [1]{[#1]}%
\providecommand \BibitemOpen [0]{}%
\providecommand \bibitemStop [0]{}%
\providecommand \bibitemNoStop [0]{.\EOS\space}%
\providecommand \EOS [0]{\spacefactor3000\relax}%
\providecommand \BibitemShut  [1]{\csname bibitem#1\endcsname}%
\let\auto@bib@innerbib\@empty
\bibitem [{\citenamefont {Ginzburg}\ and\ \citenamefont
  {Ptuskin}(1976)}]{Ginzburg:1976dj}%
  \BibitemOpen
  \bibfield  {author} {\bibinfo {author} {\bibfnamefont {V.~L.}\ \bibnamefont
  {Ginzburg}}\ and\ \bibinfo {author} {\bibfnamefont {V.~S.}\ \bibnamefont
  {Ptuskin}},\ }\href {https://doi.org/10.1103/RevModPhys.48.161} {\bibfield
  {journal} {\bibinfo  {journal} {Rev. Mod. Phys.}\ }\textbf {\bibinfo {volume}
  {48}},\ \bibinfo {pages} {161} (\bibinfo {year} {1976})},\ \bibinfo {note}
  {[Erratum: Rev.Mod.Phys. 48, 675--676 (1976)]}\BibitemShut {NoStop}%
\bibitem [{\citenamefont {Trotta}\ \emph {et~al.}(2011)\citenamefont {Trotta},
  \citenamefont {J\'ohannesson}, \citenamefont {Moskalenko}, \citenamefont
  {Porter}, \citenamefont {Austri},\ and\ \citenamefont
  {Strong}}]{Trotta:2010mx}%
  \BibitemOpen
  \bibfield  {author} {\bibinfo {author} {\bibfnamefont {R.}~\bibnamefont
  {Trotta}}, \bibinfo {author} {\bibfnamefont {G.}~\bibnamefont
  {J\'ohannesson}}, \bibinfo {author} {\bibfnamefont {I.~V.}\ \bibnamefont
  {Moskalenko}}, \bibinfo {author} {\bibfnamefont {T.~A.}\ \bibnamefont
  {Porter}}, \bibinfo {author} {\bibfnamefont {R.~R.~d.}\ \bibnamefont
  {Austri}},\ and\ \bibinfo {author} {\bibfnamefont {A.~W.}\ \bibnamefont
  {Strong}},\ }\href {https://doi.org/10.1088/0004-637X/729/2/106} {\bibfield
  {journal} {\bibinfo  {journal} {Astrophys. J.}\ }\textbf {\bibinfo {volume}
  {729}},\ \bibinfo {pages} {106} (\bibinfo {year} {2011})},\ \Eprint
  {https://arxiv.org/abs/1011.0037} {arXiv:1011.0037 [astro-ph.HE]}
  \BibitemShut {NoStop}%
\bibitem [{\citenamefont {G\'enolini}\ \emph {et~al.}(2019)\citenamefont
  {G\'enolini} \emph {et~al.}}]{Genolini:2019ewc}%
  \BibitemOpen
  \bibfield  {author} {\bibinfo {author} {\bibfnamefont {Y.}~\bibnamefont
  {G\'enolini}} \emph {et~al.},\ }\href
  {https://doi.org/10.1103/PhysRevD.99.123028} {\bibfield  {journal} {\bibinfo
  {journal} {Phys. Rev. D}\ }\textbf {\bibinfo {volume} {99}},\ \bibinfo
  {pages} {123028} (\bibinfo {year} {2019})},\ \Eprint
  {https://arxiv.org/abs/1904.08917} {arXiv:1904.08917 [astro-ph.HE]}
  \BibitemShut {NoStop}%
\bibitem [{\citenamefont {Yuan}(2019)}]{Yuan:2018vgk}%
  \BibitemOpen
  \bibfield  {author} {\bibinfo {author} {\bibfnamefont {Q.}~\bibnamefont
  {Yuan}},\ }\href {https://doi.org/10.1007/s11433-018-9300-0} {\bibfield
  {journal} {\bibinfo  {journal} {Sci. China Phys. Mech. Astron.}\ }\textbf
  {\bibinfo {volume} {62}},\ \bibinfo {pages} {49511} (\bibinfo {year}
  {2019})},\ \Eprint {https://arxiv.org/abs/1805.10649} {arXiv:1805.10649
  [astro-ph.HE]} \BibitemShut {NoStop}%
\bibitem [{\citenamefont {Adriani}\ \emph {et~al.}(2011)\citenamefont {Adriani}
  \emph {et~al.}}]{Adriani:2011cu}%
  \BibitemOpen
  \bibfield  {author} {\bibinfo {author} {\bibfnamefont {O.}~\bibnamefont
  {Adriani}} \emph {et~al.} (\bibinfo {collaboration} {PAMELA}),\ }\href
  {https://doi.org/10.1126/science.1199172} {\bibfield  {journal} {\bibinfo
  {journal} {Science}\ }\textbf {\bibinfo {volume} {332}},\ \bibinfo {pages}
  {69} (\bibinfo {year} {2011})},\ \Eprint {https://arxiv.org/abs/1103.4055}
  {arXiv:1103.4055 [astro-ph.HE]} \BibitemShut {NoStop}%
\bibitem [{\citenamefont {Aguilar}\ \emph {et~al.}(2018)\citenamefont {Aguilar}
  \emph {et~al.}}]{Aguilar:2018njt}%
  \BibitemOpen
  \bibfield  {author} {\bibinfo {author} {\bibfnamefont {M.}~\bibnamefont
  {Aguilar}} \emph {et~al.} (\bibinfo {collaboration} {AMS}),\ }\href
  {https://doi.org/10.1103/PhysRevLett.120.021101} {\bibfield  {journal}
  {\bibinfo  {journal} {Phys. Rev. Lett.}\ }\textbf {\bibinfo {volume} {120}},\
  \bibinfo {pages} {021101} (\bibinfo {year} {2018})}\BibitemShut {NoStop}%
\bibitem [{\citenamefont {Consolandi}(2016)}]{Consolandi:2016fhd}%
  \BibitemOpen
  \bibfield  {author} {\bibinfo {author} {\bibfnamefont {C.}~\bibnamefont
  {Consolandi}} (\bibinfo {collaboration} {AMS}),\ }in\ \href@noop {} {\emph
  {\bibinfo {booktitle} {{25th European Cosmic Ray Symposium}}}}\ (\bibinfo
  {year} {2016})\ \Eprint {https://arxiv.org/abs/1612.08562} {arXiv:1612.08562
  [astro-ph.HE]} \BibitemShut {NoStop}%
\bibitem [{\citenamefont {Aguilar}\ \emph {et~al.}(2017)\citenamefont {Aguilar}
  \emph {et~al.}}]{Aguilar:2017hno}%
  \BibitemOpen
  \bibfield  {author} {\bibinfo {author} {\bibfnamefont {M.}~\bibnamefont
  {Aguilar}} \emph {et~al.} (\bibinfo {collaboration} {AMS}),\ }\href
  {https://doi.org/10.1103/PhysRevLett.119.251101} {\bibfield  {journal}
  {\bibinfo  {journal} {Phys. Rev. Lett.}\ }\textbf {\bibinfo {volume} {119}},\
  \bibinfo {pages} {251101} (\bibinfo {year} {2017})}\BibitemShut {NoStop}%
\bibitem [{\citenamefont {Aguilar}\ \emph
  {et~al.}(2021{\natexlab{a}})\citenamefont {Aguilar} \emph
  {et~al.}}]{Aguilar:2021tos}%
  \BibitemOpen
  \bibfield  {author} {\bibinfo {author} {\bibfnamefont {M.}~\bibnamefont
  {Aguilar}} \emph {et~al.} (\bibinfo {collaboration} {AMS}),\ }\href
  {https://doi.org/10.1016/j.physrep.2020.09.003} {\bibfield  {journal}
  {\bibinfo  {journal} {Phys. Rept.}\ }\textbf {\bibinfo {volume} {894}},\
  \bibinfo {pages} {1} (\bibinfo {year} {2021}{\natexlab{a}})}\BibitemShut
  {NoStop}%
\bibitem [{\citenamefont {Aguilar}\ \emph
  {et~al.}(2021{\natexlab{b}})\citenamefont {Aguilar} \emph
  {et~al.}}]{AMS:2021brg}%
  \BibitemOpen
  \bibfield  {author} {\bibinfo {author} {\bibfnamefont {M.}~\bibnamefont
  {Aguilar}} \emph {et~al.} (\bibinfo {collaboration} {AMS}),\ }\href
  {https://doi.org/10.1103/PhysRevLett.127.021101} {\bibfield  {journal}
  {\bibinfo  {journal} {Phys. Rev. Lett.}\ }\textbf {\bibinfo {volume} {127}},\
  \bibinfo {pages} {02101} (\bibinfo {year} {2021}{\natexlab{b}})},\ \bibinfo
  {note} {[Erratum: Phys.Rev.Lett. 127, 159901 (2021)]}\BibitemShut {NoStop}%
\bibitem [{\citenamefont {Boschini}\ \emph {et~al.}(2020)\citenamefont
  {Boschini} \emph {et~al.}}]{Boschini:2019gow}%
  \BibitemOpen
  \bibfield  {author} {\bibinfo {author} {\bibfnamefont {M.~J.}\ \bibnamefont
  {Boschini}} \emph {et~al.},\ }\href
  {https://doi.org/10.3847/1538-4357/ab64f1} {\bibfield  {journal} {\bibinfo
  {journal} {Astrophys. J.}\ }\textbf {\bibinfo {volume} {889}},\ \bibinfo
  {pages} {167} (\bibinfo {year} {2020})},\ \Eprint
  {https://arxiv.org/abs/1911.03108} {arXiv:1911.03108 [astro-ph.HE]}
  \BibitemShut {NoStop}%
\bibitem [{\citenamefont {Luque}\ \emph {et~al.}(2021)\citenamefont {Luque},
  \citenamefont {Mazziotta}, \citenamefont {Loparco}, \citenamefont {Gargano},\
  and\ \citenamefont {Serini}}]{Luque:2021nxb}%
  \BibitemOpen
  \bibfield  {author} {\bibinfo {author} {\bibfnamefont {P.~D. L.~T.}\
  \bibnamefont {Luque}}, \bibinfo {author} {\bibfnamefont {M.~N.}\ \bibnamefont
  {Mazziotta}}, \bibinfo {author} {\bibfnamefont {F.}~\bibnamefont {Loparco}},
  \bibinfo {author} {\bibfnamefont {F.}~\bibnamefont {Gargano}},\ and\ \bibinfo
  {author} {\bibfnamefont {D.}~\bibnamefont {Serini}},\ }\href
  {https://doi.org/10.1088/1475-7516/2021/07/010} {\bibfield  {journal}
  {\bibinfo  {journal} {JCAP}\ }\textbf {\bibinfo {volume} {07}},\ \bibinfo
  {pages} {010}},\ \Eprint {https://arxiv.org/abs/2102.13238} {arXiv:2102.13238
  [astro-ph.HE]} \BibitemShut {NoStop}%
\bibitem [{\citenamefont {de~la Torre~Luque}\ \emph {et~al.}(2022)\citenamefont
  {de~la Torre~Luque}, \citenamefont {Mazziotta}, \citenamefont {Ferrari},
  \citenamefont {Loparco}, \citenamefont {Sala},\ and\ \citenamefont
  {Serini}}]{delaTorreLuque:2022vhm}%
  \BibitemOpen
  \bibfield  {author} {\bibinfo {author} {\bibfnamefont {P.}~\bibnamefont
  {de~la Torre~Luque}}, \bibinfo {author} {\bibfnamefont {M.~N.}\ \bibnamefont
  {Mazziotta}}, \bibinfo {author} {\bibfnamefont {A.}~\bibnamefont {Ferrari}},
  \bibinfo {author} {\bibfnamefont {F.}~\bibnamefont {Loparco}}, \bibinfo
  {author} {\bibfnamefont {P.}~\bibnamefont {Sala}},\ and\ \bibinfo {author}
  {\bibfnamefont {D.}~\bibnamefont {Serini}},\ }\href
  {https://doi.org/10.1088/1475-7516/2022/07/008} {\bibfield  {journal}
  {\bibinfo  {journal} {JCAP}\ }\textbf {\bibinfo {volume} {07}}\bibfield
  {number} {\bibinfo  {number} { (07)},\ \bibinfo {pages} {008}},\ }\Eprint
  {https://arxiv.org/abs/2202.03559} {arXiv:2202.03559 [astro-ph.HE]}
  \BibitemShut {NoStop}%
\bibitem [{\citenamefont {Weinrich}\ \emph
  {et~al.}(2020{\natexlab{a}})\citenamefont {Weinrich}, \citenamefont
  {G\'enolini}, \citenamefont {Boudaud}, \citenamefont {Derome},\ and\
  \citenamefont {Maurin}}]{Weinrich:2020cmw}%
  \BibitemOpen
  \bibfield  {author} {\bibinfo {author} {\bibfnamefont {N.}~\bibnamefont
  {Weinrich}}, \bibinfo {author} {\bibfnamefont {Y.}~\bibnamefont
  {G\'enolini}}, \bibinfo {author} {\bibfnamefont {M.}~\bibnamefont {Boudaud}},
  \bibinfo {author} {\bibfnamefont {L.}~\bibnamefont {Derome}},\ and\ \bibinfo
  {author} {\bibfnamefont {D.}~\bibnamefont {Maurin}},\ }\href
  {https://doi.org/10.1051/0004-6361/202037875} {\bibfield  {journal} {\bibinfo
   {journal} {Astron. Astrophys.}\ }\textbf {\bibinfo {volume} {639}},\
  \bibinfo {pages} {A131} (\bibinfo {year} {2020}{\natexlab{a}})},\ \Eprint
  {https://arxiv.org/abs/2002.11406} {arXiv:2002.11406 [astro-ph.HE]}
  \BibitemShut {NoStop}%
\bibitem [{\citenamefont {Korsmeier}\ and\ \citenamefont
  {Cuoco}(2021)}]{Korsmeier:2021brc}%
  \BibitemOpen
  \bibfield  {author} {\bibinfo {author} {\bibfnamefont {M.}~\bibnamefont
  {Korsmeier}}\ and\ \bibinfo {author} {\bibfnamefont {A.}~\bibnamefont
  {Cuoco}},\ }\href {https://doi.org/10.1103/PhysRevD.103.103016} {\bibfield
  {journal} {\bibinfo  {journal} {Phys. Rev. D}\ }\textbf {\bibinfo {volume}
  {103}},\ \bibinfo {pages} {103016} (\bibinfo {year} {2021})},\ \Eprint
  {https://arxiv.org/abs/2103.09824} {arXiv:2103.09824 [astro-ph.HE]}
  \BibitemShut {NoStop}%
\bibitem [{\citenamefont {Kawanaka}\ and\ \citenamefont
  {Yanagita}(2018)}]{Kawanaka:2017cae}%
  \BibitemOpen
  \bibfield  {author} {\bibinfo {author} {\bibfnamefont {N.}~\bibnamefont
  {Kawanaka}}\ and\ \bibinfo {author} {\bibfnamefont {S.}~\bibnamefont
  {Yanagita}},\ }\href {https://doi.org/10.1103/PhysRevLett.120.041103}
  {\bibfield  {journal} {\bibinfo  {journal} {Phys. Rev. Lett.}\ }\textbf
  {\bibinfo {volume} {120}},\ \bibinfo {pages} {041103} (\bibinfo {year}
  {2018})},\ \Eprint {https://arxiv.org/abs/1707.00212} {arXiv:1707.00212
  [astro-ph.HE]} \BibitemShut {NoStop}%
\bibitem [{\citenamefont {Maurin}\ \emph {et~al.}(2022)\citenamefont {Maurin},
  \citenamefont {Bueno}, \citenamefont {G\'enolini}, \citenamefont {Derome},\
  and\ \citenamefont {Vecchi}}]{Maurin:2022irz}%
  \BibitemOpen
  \bibfield  {author} {\bibinfo {author} {\bibfnamefont {D.}~\bibnamefont
  {Maurin}}, \bibinfo {author} {\bibfnamefont {E.~F.}\ \bibnamefont {Bueno}},
  \bibinfo {author} {\bibfnamefont {Y.}~\bibnamefont {G\'enolini}}, \bibinfo
  {author} {\bibfnamefont {L.}~\bibnamefont {Derome}},\ and\ \bibinfo {author}
  {\bibfnamefont {M.}~\bibnamefont {Vecchi}},\ }\href
  {https://doi.org/10.1051/0004-6361/202243446} {\bibfield  {journal} {\bibinfo
   {journal} {Astron. Astrophys.}\ }\textbf {\bibinfo {volume} {668}},\
  \bibinfo {pages} {A7} (\bibinfo {year} {2022})},\ \Eprint
  {https://arxiv.org/abs/2203.00522} {arXiv:2203.00522 [astro-ph.HE]}
  \BibitemShut {NoStop}%
\bibitem [{\citenamefont {Zhao}\ \emph {et~al.}(2024)\citenamefont {Zhao},
  \citenamefont {Bi}, \citenamefont {Fang},\ and\ \citenamefont
  {Yin}}]{Zhao:2024qbj}%
  \BibitemOpen
  \bibfield  {author} {\bibinfo {author} {\bibfnamefont {M.-J.}\ \bibnamefont
  {Zhao}}, \bibinfo {author} {\bibfnamefont {X.-J.}\ \bibnamefont {Bi}},
  \bibinfo {author} {\bibfnamefont {K.}~\bibnamefont {Fang}},\ and\ \bibinfo
  {author} {\bibfnamefont {P.-F.}\ \bibnamefont {Yin}},\ }\href
  {https://doi.org/10.1103/PhysRevD.109.083036} {\bibfield  {journal} {\bibinfo
   {journal} {Phys. Rev. D}\ }\textbf {\bibinfo {volume} {109}},\ \bibinfo
  {pages} {083036} (\bibinfo {year} {2024})},\ \Eprint
  {https://arxiv.org/abs/2402.04659} {arXiv:2402.04659 [astro-ph.HE]}
  \BibitemShut {NoStop}%
\bibitem [{\citenamefont {Derome}(2021)}]{AMS_icrc2021}%
  \BibitemOpen
  \bibfield  {author} {\bibinfo {author} {\bibfnamefont {L.}~\bibnamefont
  {Derome}},\ }\href@noop {} {\bibinfo {title} {Cosmic-ray lithium and
  beryllium isotopes with ams02}},\ \bibinfo {howpublished}
  {\url{https://indico.desy.de/event/27991/contributions/101805/}} (\bibinfo
  {year} {2021})\BibitemShut {NoStop}%
\bibitem [{\citenamefont {Wei}(2022)}]{AMS_ichep}%
  \BibitemOpen
  \bibfield  {author} {\bibinfo {author} {\bibfnamefont {J.}~\bibnamefont
  {Wei}},\ }\href@noop {} {\bibinfo {title} {Properties of cosmic beryllium
  isotopes}},\ \bibinfo {howpublished}
  {\url{https://agenda.infn.it/event/28874/contributions/170166/}} (\bibinfo
  {year} {2022})\BibitemShut {NoStop}%
\bibitem [{\citenamefont {Lv}\ \emph {et~al.}(2024)\citenamefont {Lv},
  \citenamefont {Bi}, \citenamefont {Fang}, \citenamefont {Yin},\ and\
  \citenamefont {Zhao}}]{Lv:2024tls}%
  \BibitemOpen
  \bibfield  {author} {\bibinfo {author} {\bibfnamefont {X.-J.}\ \bibnamefont
  {Lv}}, \bibinfo {author} {\bibfnamefont {X.-J.}\ \bibnamefont {Bi}}, \bibinfo
  {author} {\bibfnamefont {K.}~\bibnamefont {Fang}}, \bibinfo {author}
  {\bibfnamefont {P.-F.}\ \bibnamefont {Yin}},\ and\ \bibinfo {author}
  {\bibfnamefont {M.-J.}\ \bibnamefont {Zhao}},\ }\href
  {https://doi.org/10.1103/PhysRevD.110.123030} {\bibfield  {journal} {\bibinfo
   {journal} {Phys. Rev. D}\ }\textbf {\bibinfo {volume} {110}},\ \bibinfo
  {pages} {123030} (\bibinfo {year} {2024})},\ \Eprint
  {https://arxiv.org/abs/2409.07139} {arXiv:2409.07139 [astro-ph.HE]}
  \BibitemShut {NoStop}%
\bibitem [{\citenamefont {Aguilar}\ \emph {et~al.}(2024)\citenamefont {Aguilar}
  \emph {et~al.}}]{AMS:2024idr}%
  \BibitemOpen
  \bibfield  {author} {\bibinfo {author} {\bibfnamefont {M.}~\bibnamefont
  {Aguilar}} \emph {et~al.} (\bibinfo {collaboration} {AMS}),\ }\href
  {https://doi.org/10.1103/PhysRevLett.132.261001} {\bibfield  {journal}
  {\bibinfo  {journal} {Phys. Rev. Lett.}\ }\textbf {\bibinfo {volume} {132}},\
  \bibinfo {pages} {261001} (\bibinfo {year} {2024})}\BibitemShut {NoStop}%
\bibitem [{\citenamefont {Silberberg}\ \emph {et~al.}(1998)\citenamefont
  {Silberberg}, \citenamefont {Tsao},\ and\ \citenamefont
  {Barghouty}}]{Silberberg:1998lxa}%
  \BibitemOpen
  \bibfield  {author} {\bibinfo {author} {\bibfnamefont {R.}~\bibnamefont
  {Silberberg}}, \bibinfo {author} {\bibfnamefont {C.~H.}\ \bibnamefont
  {Tsao}},\ and\ \bibinfo {author} {\bibfnamefont {A.~F.}\ \bibnamefont
  {Barghouty}},\ }\href {https://doi.org/10.1086/305862} {\bibfield  {journal}
  {\bibinfo  {journal} {Astrophys. J.}\ }\textbf {\bibinfo {volume} {501}},\
  \bibinfo {pages} {911} (\bibinfo {year} {1998})}\BibitemShut {NoStop}%
\bibitem [{\citenamefont {Strong}\ and\ \citenamefont
  {Moskalenko}(1998)}]{Strong:1998pw}%
  \BibitemOpen
  \bibfield  {author} {\bibinfo {author} {\bibfnamefont {A.~W.}\ \bibnamefont
  {Strong}}\ and\ \bibinfo {author} {\bibfnamefont {I.~V.}\ \bibnamefont
  {Moskalenko}},\ }\href {https://doi.org/10.1086/306470} {\bibfield  {journal}
  {\bibinfo  {journal} {Astrophys. J.}\ }\textbf {\bibinfo {volume} {509}},\
  \bibinfo {pages} {212} (\bibinfo {year} {1998})},\ \Eprint
  {https://arxiv.org/abs/astro-ph/9807150} {arXiv:astro-ph/9807150}
  \BibitemShut {NoStop}%
\bibitem [{\citenamefont {Strong}\ \emph {et~al.}(2000)\citenamefont {Strong},
  \citenamefont {Moskalenko},\ and\ \citenamefont {Reimer}}]{Strong:1998fr}%
  \BibitemOpen
  \bibfield  {author} {\bibinfo {author} {\bibfnamefont {A.~W.}\ \bibnamefont
  {Strong}}, \bibinfo {author} {\bibfnamefont {I.~V.}\ \bibnamefont
  {Moskalenko}},\ and\ \bibinfo {author} {\bibfnamefont {O.}~\bibnamefont
  {Reimer}},\ }\href {https://doi.org/10.1086/309038} {\bibfield  {journal}
  {\bibinfo  {journal} {Astrophys. J.}\ }\textbf {\bibinfo {volume} {537}},\
  \bibinfo {pages} {763} (\bibinfo {year} {2000})},\ \bibinfo {note} {[Erratum:
  Astrophys.J. 541, 1109 (2000)]},\ \Eprint
  {https://arxiv.org/abs/astro-ph/9811296} {arXiv:astro-ph/9811296}
  \BibitemShut {NoStop}%
\bibitem [{\citenamefont {Porter}\ \emph {et~al.}(2022)\citenamefont {Porter},
  \citenamefont {Johannesson},\ and\ \citenamefont
  {Moskalenko}}]{Porter:2021tlr}%
  \BibitemOpen
  \bibfield  {author} {\bibinfo {author} {\bibfnamefont {T.~A.}\ \bibnamefont
  {Porter}}, \bibinfo {author} {\bibfnamefont {G.}~\bibnamefont
  {Johannesson}},\ and\ \bibinfo {author} {\bibfnamefont {I.~V.}\ \bibnamefont
  {Moskalenko}},\ }\href {https://doi.org/10.3847/1538-4365/ac80f6} {\bibfield
  {journal} {\bibinfo  {journal} {Astrophys. J. Supp.}\ }\textbf {\bibinfo
  {volume} {262}},\ \bibinfo {pages} {30} (\bibinfo {year} {2022})},\ \Eprint
  {https://arxiv.org/abs/2112.12745} {arXiv:2112.12745 [astro-ph.HE]}
  \BibitemShut {NoStop}%
\bibitem [{\citenamefont {{Seo}}\ and\ \citenamefont
  {{Ptuskin}}(1994)}]{1994ApJ...431..705S}%
  \BibitemOpen
  \bibfield  {author} {\bibinfo {author} {\bibfnamefont {E.~S.}\ \bibnamefont
  {{Seo}}}\ and\ \bibinfo {author} {\bibfnamefont {V.~S.}\ \bibnamefont
  {{Ptuskin}}},\ }\href {https://doi.org/10.1086/174520} {\bibfield  {journal}
  {\bibinfo  {journal} {\apj}\ }\textbf {\bibinfo {volume} {431}},\ \bibinfo
  {pages} {705} (\bibinfo {year} {1994})}\BibitemShut {NoStop}%
\bibitem [{\citenamefont {Amin}(2021)}]{Amin:2021oow}%
  \BibitemOpen
  \bibfield  {author} {\bibinfo {author} {\bibfnamefont {N.}~\bibnamefont
  {Amin}} (\bibinfo {collaboration} {NA61/SHINE}),\ }\href
  {https://doi.org/10.22323/1.395.0102} {\bibfield  {journal} {\bibinfo
  {journal} {PoS}\ }\textbf {\bibinfo {volume} {ICRC2021}},\ \bibinfo {pages}
  {102} (\bibinfo {year} {2021})},\ \Eprint {https://arxiv.org/abs/2107.12275}
  {arXiv:2107.12275 [nucl-ex]} \BibitemShut {NoStop}%
\bibitem [{\citenamefont {Amin}(2023)}]{Amin:2023fki}%
  \BibitemOpen
  \bibfield  {author} {\bibinfo {author} {\bibfnamefont {N.}~\bibnamefont
  {Amin}} (\bibinfo {collaboration} {NA61/SHINE}),\ }\href
  {https://doi.org/10.22323/1.444.0075} {\bibfield  {journal} {\bibinfo
  {journal} {PoS}\ }\textbf {\bibinfo {volume} {ICRC2023}},\ \bibinfo {pages}
  {075} (\bibinfo {year} {2023})}\BibitemShut {NoStop}%
\bibitem [{\citenamefont {Webber}\ \emph
  {et~al.}(1990{\natexlab{a}})\citenamefont {Webber}, \citenamefont {Kish},\
  and\ \citenamefont {Schrier}}]{Webber:1990kb}%
  \BibitemOpen
  \bibfield  {author} {\bibinfo {author} {\bibfnamefont {W.~R.}\ \bibnamefont
  {Webber}}, \bibinfo {author} {\bibfnamefont {J.~C.}\ \bibnamefont {Kish}},\
  and\ \bibinfo {author} {\bibfnamefont {D.~A.}\ \bibnamefont {Schrier}},\
  }\href {https://doi.org/10.1103/PhysRevC.41.533} {\bibfield  {journal}
  {\bibinfo  {journal} {Phys. Rev. C}\ }\textbf {\bibinfo {volume} {41}},\
  \bibinfo {pages} {533} (\bibinfo {year} {1990}{\natexlab{a}})}\BibitemShut
  {NoStop}%
\bibitem [{\citenamefont {Zeitlin}\ \emph {et~al.}(2001)\citenamefont
  {Zeitlin}, \citenamefont {Fukumura}, \citenamefont {Heilbronn}, \citenamefont
  {Iwata}, \citenamefont {Miller},\ and\ \citenamefont
  {Murakami}}]{Zeitlin:2001ye}%
  \BibitemOpen
  \bibfield  {author} {\bibinfo {author} {\bibfnamefont {C.}~\bibnamefont
  {Zeitlin}}, \bibinfo {author} {\bibfnamefont {A.}~\bibnamefont {Fukumura}},
  \bibinfo {author} {\bibfnamefont {L.}~\bibnamefont {Heilbronn}}, \bibinfo
  {author} {\bibfnamefont {Y.}~\bibnamefont {Iwata}}, \bibinfo {author}
  {\bibfnamefont {J.}~\bibnamefont {Miller}},\ and\ \bibinfo {author}
  {\bibfnamefont {T.}~\bibnamefont {Murakami}},\ }\href
  {https://doi.org/10.1103/PhysRevC.64.024902} {\bibfield  {journal} {\bibinfo
  {journal} {Phys. Rev. C}\ }\textbf {\bibinfo {volume} {64}},\ \bibinfo
  {pages} {024902} (\bibinfo {year} {2001})}\BibitemShut {NoStop}%
\bibitem [{\citenamefont {Zeitlin}\ \emph {et~al.}(2007)\citenamefont
  {Zeitlin}, \citenamefont {Guetersloh}, \citenamefont {Heilbronn},
  \citenamefont {Fukumura}, \citenamefont {Iwata}, \citenamefont {Miller},\
  and\ \citenamefont {Murakami}}]{Zeitlin:2007sm}%
  \BibitemOpen
  \bibfield  {author} {\bibinfo {author} {\bibfnamefont {C.}~\bibnamefont
  {Zeitlin}}, \bibinfo {author} {\bibfnamefont {S.~B.}\ \bibnamefont
  {Guetersloh}}, \bibinfo {author} {\bibfnamefont {L.~H.}\ \bibnamefont
  {Heilbronn}}, \bibinfo {author} {\bibfnamefont {A.}~\bibnamefont {Fukumura}},
  \bibinfo {author} {\bibfnamefont {Y.}~\bibnamefont {Iwata}}, \bibinfo
  {author} {\bibfnamefont {J.}~\bibnamefont {Miller}},\ and\ \bibinfo {author}
  {\bibfnamefont {T.}~\bibnamefont {Murakami}},\ }\href
  {https://doi.org/10.1016/j.nuclphysa.2006.10.088} {\bibfield  {journal}
  {\bibinfo  {journal} {Nucl. Phys. A}\ }\textbf {\bibinfo {volume} {784}},\
  \bibinfo {pages} {341} (\bibinfo {year} {2007})}\BibitemShut {NoStop}%
\bibitem [{\citenamefont {Zeitlin}\ \emph {et~al.}(2011)\citenamefont
  {Zeitlin}, \citenamefont {Miller}, \citenamefont {Guetersloh}, \citenamefont
  {Heilbronn}, \citenamefont {Fukumura}, \citenamefont {Iwata}, \citenamefont
  {Murakami}, \citenamefont {Blattnig}, \citenamefont {Norman},\ and\
  \citenamefont {Mashnik}}]{Zeitlin:2011qg}%
  \BibitemOpen
  \bibfield  {author} {\bibinfo {author} {\bibfnamefont {C.}~\bibnamefont
  {Zeitlin}}, \bibinfo {author} {\bibfnamefont {J.}~\bibnamefont {Miller}},
  \bibinfo {author} {\bibfnamefont {S.}~\bibnamefont {Guetersloh}}, \bibinfo
  {author} {\bibfnamefont {L.}~\bibnamefont {Heilbronn}}, \bibinfo {author}
  {\bibfnamefont {A.}~\bibnamefont {Fukumura}}, \bibinfo {author}
  {\bibfnamefont {Y.}~\bibnamefont {Iwata}}, \bibinfo {author} {\bibfnamefont
  {T.}~\bibnamefont {Murakami}}, \bibinfo {author} {\bibfnamefont
  {S.}~\bibnamefont {Blattnig}}, \bibinfo {author} {\bibfnamefont
  {R.}~\bibnamefont {Norman}},\ and\ \bibinfo {author} {\bibfnamefont
  {S.}~\bibnamefont {Mashnik}},\ }\href
  {https://doi.org/10.1103/PhysRevC.83.034909} {\bibfield  {journal} {\bibinfo
  {journal} {Phys. Rev. C}\ }\textbf {\bibinfo {volume} {83}},\ \bibinfo
  {pages} {034909} (\bibinfo {year} {2011})},\ \Eprint
  {https://arxiv.org/abs/1102.2848} {arXiv:1102.2848 [nucl-ex]} \BibitemShut
  {NoStop}%
\bibitem [{\citenamefont {J\'ohannesson}\ \emph {et~al.}(2016)\citenamefont
  {J\'ohannesson} \emph {et~al.}}]{Johannesson:2016rlh}%
  \BibitemOpen
  \bibfield  {author} {\bibinfo {author} {\bibfnamefont {G.}~\bibnamefont
  {J\'ohannesson}} \emph {et~al.},\ }\href
  {https://doi.org/10.3847/0004-637X/824/1/16} {\bibfield  {journal} {\bibinfo
  {journal} {Astrophys. J.}\ }\textbf {\bibinfo {volume} {824}},\ \bibinfo
  {pages} {16} (\bibinfo {year} {2016})},\ \Eprint
  {https://arxiv.org/abs/1602.02243} {arXiv:1602.02243 [astro-ph.HE]}
  \BibitemShut {NoStop}%
\bibitem [{\citenamefont {Phan}\ \emph {et~al.}(2021)\citenamefont {Phan},
  \citenamefont {Schulze}, \citenamefont {Mertsch}, \citenamefont {Recchia},\
  and\ \citenamefont {Gabici}}]{Phan:2021iht}%
  \BibitemOpen
  \bibfield  {author} {\bibinfo {author} {\bibfnamefont {V.~H.~M.}\
  \bibnamefont {Phan}}, \bibinfo {author} {\bibfnamefont {F.}~\bibnamefont
  {Schulze}}, \bibinfo {author} {\bibfnamefont {P.}~\bibnamefont {Mertsch}},
  \bibinfo {author} {\bibfnamefont {S.}~\bibnamefont {Recchia}},\ and\ \bibinfo
  {author} {\bibfnamefont {S.}~\bibnamefont {Gabici}},\ }\href
  {https://doi.org/10.1103/PhysRevLett.127.141101} {\bibfield  {journal}
  {\bibinfo  {journal} {Phys. Rev. Lett.}\ }\textbf {\bibinfo {volume} {127}},\
  \bibinfo {pages} {141101} (\bibinfo {year} {2021})},\ \Eprint
  {https://arxiv.org/abs/2105.00311} {arXiv:2105.00311 [astro-ph.HE]}
  \BibitemShut {NoStop}%
\bibitem [{\citenamefont {Aguilar}\ \emph {et~al.}(2020)\citenamefont {Aguilar}
  \emph {et~al.}}]{AMS:2020cai}%
  \BibitemOpen
  \bibfield  {author} {\bibinfo {author} {\bibfnamefont {M.}~\bibnamefont
  {Aguilar}} \emph {et~al.} (\bibinfo {collaboration} {AMS}),\ }\href
  {https://doi.org/10.1103/PhysRevLett.124.211102} {\bibfield  {journal}
  {\bibinfo  {journal} {Phys. Rev. Lett.}\ }\textbf {\bibinfo {volume} {124}},\
  \bibinfo {pages} {211102} (\bibinfo {year} {2020})}\BibitemShut {NoStop}%
\bibitem [{\citenamefont {Aguilar}\ \emph
  {et~al.}(2021{\natexlab{c}})\citenamefont {Aguilar} \emph
  {et~al.}}]{AMS:2021lxc}%
  \BibitemOpen
  \bibfield  {author} {\bibinfo {author} {\bibfnamefont {M.}~\bibnamefont
  {Aguilar}} \emph {et~al.} (\bibinfo {collaboration} {AMS}),\ }\href
  {https://doi.org/10.1103/PhysRevLett.126.041104} {\bibfield  {journal}
  {\bibinfo  {journal} {Phys. Rev. Lett.}\ }\textbf {\bibinfo {volume} {126}},\
  \bibinfo {pages} {041104} (\bibinfo {year} {2021}{\natexlab{c}})}\BibitemShut
  {NoStop}%
\bibitem [{\citenamefont {Aguilar}\ \emph {et~al.}(2023)\citenamefont {Aguilar}
  \emph {et~al.}}]{AMS:2023anq}%
  \BibitemOpen
  \bibfield  {author} {\bibinfo {author} {\bibfnamefont {M.}~\bibnamefont
  {Aguilar}} \emph {et~al.} (\bibinfo {collaboration} {AMS}),\ }\href
  {https://doi.org/10.1103/PhysRevLett.130.211002} {\bibfield  {journal}
  {\bibinfo  {journal} {Phys. Rev. Lett.}\ }\textbf {\bibinfo {volume} {130}},\
  \bibinfo {pages} {211002} (\bibinfo {year} {2023})}\BibitemShut {NoStop}%
\bibitem [{\citenamefont {Cummings}\ \emph {et~al.}(2016)\citenamefont
  {Cummings}, \citenamefont {Stone}, \citenamefont {Heikkila}, \citenamefont
  {Lal}, \citenamefont {Webber}, \citenamefont {J\'ohannesson}, \citenamefont
  {Moskalenko}, \citenamefont {Orlando},\ and\ \citenamefont
  {Porter}}]{Cummings:2016pdr}%
  \BibitemOpen
  \bibfield  {author} {\bibinfo {author} {\bibfnamefont {A.~C.}\ \bibnamefont
  {Cummings}}, \bibinfo {author} {\bibfnamefont {E.~C.}\ \bibnamefont {Stone}},
  \bibinfo {author} {\bibfnamefont {B.~C.}\ \bibnamefont {Heikkila}}, \bibinfo
  {author} {\bibfnamefont {N.}~\bibnamefont {Lal}}, \bibinfo {author}
  {\bibfnamefont {W.~R.}\ \bibnamefont {Webber}}, \bibinfo {author}
  {\bibfnamefont {G.}~\bibnamefont {J\'ohannesson}}, \bibinfo {author}
  {\bibfnamefont {I.~V.}\ \bibnamefont {Moskalenko}}, \bibinfo {author}
  {\bibfnamefont {E.}~\bibnamefont {Orlando}},\ and\ \bibinfo {author}
  {\bibfnamefont {T.~A.}\ \bibnamefont {Porter}},\ }\href
  {https://doi.org/10.3847/0004-637X/831/1/18} {\bibfield  {journal} {\bibinfo
  {journal} {Astrophys. J.}\ }\textbf {\bibinfo {volume} {831}},\ \bibinfo
  {pages} {18} (\bibinfo {year} {2016})}\BibitemShut {NoStop}%
\bibitem [{\citenamefont {Boschini}\ \emph {et~al.}(2022)\citenamefont
  {Boschini} \emph {et~al.}}]{Boschini:2022fpd}%
  \BibitemOpen
  \bibfield  {author} {\bibinfo {author} {\bibfnamefont {M.~J.}\ \bibnamefont
  {Boschini}} \emph {et~al.},\ }\href
  {https://doi.org/10.3847/1538-4357/ac7443} {\bibfield  {journal} {\bibinfo
  {journal} {Astrophys. J.}\ }\textbf {\bibinfo {volume} {933}},\ \bibinfo
  {pages} {147} (\bibinfo {year} {2022})},\ \Eprint
  {https://arxiv.org/abs/2202.09928} {arXiv:2202.09928 [astro-ph.HE]}
  \BibitemShut {NoStop}%
\bibitem [{\citenamefont {Gleeson}\ and\ \citenamefont
  {Axford}(1968)}]{Gleeson:1968zza}%
  \BibitemOpen
  \bibfield  {author} {\bibinfo {author} {\bibfnamefont {L.~J.}\ \bibnamefont
  {Gleeson}}\ and\ \bibinfo {author} {\bibfnamefont {W.~I.}\ \bibnamefont
  {Axford}},\ }\href {https://doi.org/10.1086/149822} {\bibfield  {journal}
  {\bibinfo  {journal} {Astrophys. J.}\ }\textbf {\bibinfo {volume} {154}},\
  \bibinfo {pages} {1011} (\bibinfo {year} {1968})}\BibitemShut {NoStop}%
\bibitem [{\citenamefont {Webber}\ \emph {et~al.}(2018)\citenamefont {Webber},
  \citenamefont {Lal},\ and\ \citenamefont
  {Heikkila}}]{webber2018measurementsinterstellarspacegalactic}%
  \BibitemOpen
  \bibfield  {author} {\bibinfo {author} {\bibfnamefont {W.~R.}\ \bibnamefont
  {Webber}}, \bibinfo {author} {\bibfnamefont {N.}~\bibnamefont {Lal}},\ and\
  \bibinfo {author} {\bibfnamefont {B.}~\bibnamefont {Heikkila}},\ }\href
  {https://arxiv.org/abs/1810.08589} {\bibinfo {title} {Measurements in
  interstellar space of galactic cosmic ray isotopes of li, be, b and n, ne
  nuclei between 40-160 mev/nuc by the crs instrument on voyager 1}} (\bibinfo
  {year} {2018}),\ \Eprint {https://arxiv.org/abs/1810.08589} {arXiv:1810.08589
  [physics.space-ph]} \BibitemShut {NoStop}%
\bibitem [{\citenamefont {Webber}\ \emph
  {et~al.}(1990{\natexlab{b}})\citenamefont {Webber}, \citenamefont {Kish},\
  and\ \citenamefont {Schrier}}]{Webber:1990pr}%
  \BibitemOpen
  \bibfield  {author} {\bibinfo {author} {\bibfnamefont {W.~R.}\ \bibnamefont
  {Webber}}, \bibinfo {author} {\bibfnamefont {J.~C.}\ \bibnamefont {Kish}},\
  and\ \bibinfo {author} {\bibfnamefont {D.~A.}\ \bibnamefont {Schrier}},\
  }\href {https://doi.org/10.1103/PhysRevC.41.566} {\bibfield  {journal}
  {\bibinfo  {journal} {Phys. Rev. C}\ }\textbf {\bibinfo {volume} {41}},\
  \bibinfo {pages} {566} (\bibinfo {year} {1990}{\natexlab{b}})}\BibitemShut
  {NoStop}%
\bibitem [{\citenamefont {Weinrich}\ \emph
  {et~al.}(2020{\natexlab{b}})\citenamefont {Weinrich}, \citenamefont
  {Boudaud}, \citenamefont {Derome}, \citenamefont {Genolini}, \citenamefont
  {Lavalle}, \citenamefont {Maurin}, \citenamefont {Salati}, \citenamefont
  {Serpico},\ and\ \citenamefont {Weymann-Despres}}]{Weinrich:2020ftb}%
  \BibitemOpen
  \bibfield  {author} {\bibinfo {author} {\bibfnamefont {N.}~\bibnamefont
  {Weinrich}}, \bibinfo {author} {\bibfnamefont {M.}~\bibnamefont {Boudaud}},
  \bibinfo {author} {\bibfnamefont {L.}~\bibnamefont {Derome}}, \bibinfo
  {author} {\bibfnamefont {Y.}~\bibnamefont {Genolini}}, \bibinfo {author}
  {\bibfnamefont {J.}~\bibnamefont {Lavalle}}, \bibinfo {author} {\bibfnamefont
  {D.}~\bibnamefont {Maurin}}, \bibinfo {author} {\bibfnamefont
  {P.}~\bibnamefont {Salati}}, \bibinfo {author} {\bibfnamefont
  {P.}~\bibnamefont {Serpico}},\ and\ \bibinfo {author} {\bibfnamefont
  {G.}~\bibnamefont {Weymann-Despres}},\ }\href
  {https://doi.org/10.1051/0004-6361/202038064} {\bibfield  {journal} {\bibinfo
   {journal} {Astron. Astrophys.}\ }\textbf {\bibinfo {volume} {639}},\
  \bibinfo {pages} {A74} (\bibinfo {year} {2020}{\natexlab{b}})},\ \Eprint
  {https://arxiv.org/abs/2004.00441} {arXiv:2004.00441 [astro-ph.HE]}
  \BibitemShut {NoStop}%
\bibitem [{\citenamefont {De~La Torre~Luque}\ \emph {et~al.}(2021)\citenamefont
  {De~La Torre~Luque}, \citenamefont {Mazziotta}, \citenamefont {Loparco},
  \citenamefont {Gargano},\ and\ \citenamefont
  {Serini}}]{DeLaTorreLuque:2021yfq}%
  \BibitemOpen
  \bibfield  {author} {\bibinfo {author} {\bibfnamefont {P.}~\bibnamefont
  {De~La Torre~Luque}}, \bibinfo {author} {\bibfnamefont {M.~N.}\ \bibnamefont
  {Mazziotta}}, \bibinfo {author} {\bibfnamefont {F.}~\bibnamefont {Loparco}},
  \bibinfo {author} {\bibfnamefont {F.}~\bibnamefont {Gargano}},\ and\ \bibinfo
  {author} {\bibfnamefont {D.}~\bibnamefont {Serini}},\ }\href
  {https://doi.org/10.1088/1475-7516/2021/03/099} {\bibfield  {journal}
  {\bibinfo  {journal} {JCAP}\ }\textbf {\bibinfo {volume} {03}},\ \bibinfo
  {pages} {099}},\ \Eprint {https://arxiv.org/abs/2101.01547} {arXiv:2101.01547
  [astro-ph.HE]} \BibitemShut {NoStop}%
\bibitem [{\citenamefont {Herbach}\ \emph {et~al.}(2006)\citenamefont {Herbach}
  \emph {et~al.}}]{Herbach:2006rw}%
  \BibitemOpen
  \bibfield  {author} {\bibinfo {author} {\bibfnamefont {C.~M.}\ \bibnamefont
  {Herbach}} \emph {et~al.},\ }\href
  {https://doi.org/10.1016/j.nuclphysa.2005.10.014} {\bibfield  {journal}
  {\bibinfo  {journal} {Nucl. Phys. A}\ }\textbf {\bibinfo {volume} {765}},\
  \bibinfo {pages} {426} (\bibinfo {year} {2006})}\BibitemShut {NoStop}%
\bibitem [{\citenamefont {Ramaty}\ \emph {et~al.}(1997)\citenamefont {Ramaty},
  \citenamefont {Kozlovsky},\ and\ \citenamefont
  {Lingenfelter}}]{Ramaty:1996qt}%
  \BibitemOpen
  \bibfield  {author} {\bibinfo {author} {\bibfnamefont {R.}~\bibnamefont
  {Ramaty}}, \bibinfo {author} {\bibfnamefont {B.}~\bibnamefont {Kozlovsky}},\
  and\ \bibinfo {author} {\bibfnamefont {R.~E.}\ \bibnamefont {Lingenfelter}},\
  }\href {https://doi.org/10.1086/304744} {\bibfield  {journal} {\bibinfo
  {journal} {Astrophys. J.}\ }\textbf {\bibinfo {volume} {488}},\ \bibinfo
  {pages} {730} (\bibinfo {year} {1997})},\ \Eprint
  {https://arxiv.org/abs/astro-ph/9610255} {arXiv:astro-ph/9610255}
  \BibitemShut {NoStop}%
\bibitem [{\citenamefont {Evoli}\ \emph {et~al.}(2018)\citenamefont {Evoli},
  \citenamefont {Gaggero}, \citenamefont {Vittino}, \citenamefont {Di~Mauro},
  \citenamefont {Grasso},\ and\ \citenamefont {Mazziotta}}]{Evoli:2017vim}%
  \BibitemOpen
  \bibfield  {author} {\bibinfo {author} {\bibfnamefont {C.}~\bibnamefont
  {Evoli}}, \bibinfo {author} {\bibfnamefont {D.}~\bibnamefont {Gaggero}},
  \bibinfo {author} {\bibfnamefont {A.}~\bibnamefont {Vittino}}, \bibinfo
  {author} {\bibfnamefont {M.}~\bibnamefont {Di~Mauro}}, \bibinfo {author}
  {\bibfnamefont {D.}~\bibnamefont {Grasso}},\ and\ \bibinfo {author}
  {\bibfnamefont {M.~N.}\ \bibnamefont {Mazziotta}},\ }\href
  {https://doi.org/10.1088/1475-7516/2018/07/006} {\bibfield  {journal}
  {\bibinfo  {journal} {JCAP}\ }\textbf {\bibinfo {volume} {07}},\ \bibinfo
  {pages} {006}},\ \Eprint {https://arxiv.org/abs/1711.09616} {arXiv:1711.09616
  [astro-ph.HE]} \BibitemShut {NoStop}%
\bibitem [{\citenamefont {Zhao}\ \emph {et~al.}(2023)\citenamefont {Zhao},
  \citenamefont {Bi},\ and\ \citenamefont {Fang}}]{Zhao:2022bon}%
  \BibitemOpen
  \bibfield  {author} {\bibinfo {author} {\bibfnamefont {M.-J.}\ \bibnamefont
  {Zhao}}, \bibinfo {author} {\bibfnamefont {X.-J.}\ \bibnamefont {Bi}},\ and\
  \bibinfo {author} {\bibfnamefont {K.}~\bibnamefont {Fang}},\ }\href
  {https://doi.org/10.1103/PhysRevD.107.063020} {\bibfield  {journal} {\bibinfo
   {journal} {Phys. Rev. D}\ }\textbf {\bibinfo {volume} {107}},\ \bibinfo
  {pages} {063020} (\bibinfo {year} {2023})},\ \Eprint
  {https://arxiv.org/abs/2209.03799} {arXiv:2209.03799 [astro-ph.HE]}
  \BibitemShut {NoStop}%
\bibitem [{\citenamefont {Silberberg}\ and\ \citenamefont
  {Tsao}(1973{\natexlab{a}})}]{Silberberg:1973jxa}%
  \BibitemOpen
  \bibfield  {author} {\bibinfo {author} {\bibfnamefont {R.}~\bibnamefont
  {Silberberg}}\ and\ \bibinfo {author} {\bibfnamefont {C.~H.}\ \bibnamefont
  {Tsao}},\ }\href {https://doi.org/10.1086/190271} {\bibfield  {journal}
  {\bibinfo  {journal} {Astrophys. J. Suppl.}\ }\textbf {\bibinfo {volume}
  {25}},\ \bibinfo {pages} {315} (\bibinfo {year}
  {1973}{\natexlab{a}})}\BibitemShut {NoStop}%
\bibitem [{\citenamefont {Genolini}\ \emph {et~al.}(2018)\citenamefont
  {Genolini}, \citenamefont {Maurin}, \citenamefont {Moskalenko},\ and\
  \citenamefont {Unger}}]{Genolini:2018ekk}%
  \BibitemOpen
  \bibfield  {author} {\bibinfo {author} {\bibfnamefont {Y.}~\bibnamefont
  {Genolini}}, \bibinfo {author} {\bibfnamefont {D.}~\bibnamefont {Maurin}},
  \bibinfo {author} {\bibfnamefont {I.~V.}\ \bibnamefont {Moskalenko}},\ and\
  \bibinfo {author} {\bibfnamefont {M.}~\bibnamefont {Unger}},\ }\href
  {https://doi.org/10.1103/PhysRevC.98.034611} {\bibfield  {journal} {\bibinfo
  {journal} {Phys. Rev. C}\ }\textbf {\bibinfo {volume} {98}},\ \bibinfo
  {pages} {034611} (\bibinfo {year} {2018})},\ \Eprint
  {https://arxiv.org/abs/1803.04686} {arXiv:1803.04686 [astro-ph.HE]}
  \BibitemShut {NoStop}%
\bibitem [{\citenamefont {Bueno}\ \emph {et~al.}(2024)\citenamefont {Bueno},
  \citenamefont {Derome}, \citenamefont {G\'enolini}, \citenamefont {Maurin},
  \citenamefont {Tatischeff},\ and\ \citenamefont {Vecchi}}]{Bueno:2022bdc}%
  \BibitemOpen
  \bibfield  {author} {\bibinfo {author} {\bibfnamefont {E.~F.}\ \bibnamefont
  {Bueno}}, \bibinfo {author} {\bibfnamefont {L.}~\bibnamefont {Derome}},
  \bibinfo {author} {\bibfnamefont {Y.}~\bibnamefont {G\'enolini}}, \bibinfo
  {author} {\bibfnamefont {D.}~\bibnamefont {Maurin}}, \bibinfo {author}
  {\bibfnamefont {V.}~\bibnamefont {Tatischeff}},\ and\ \bibinfo {author}
  {\bibfnamefont {M.}~\bibnamefont {Vecchi}},\ }\href
  {https://doi.org/10.1051/0004-6361/202244660} {\bibfield  {journal} {\bibinfo
   {journal} {Astron. Astrophys.}\ }\textbf {\bibinfo {volume} {688}},\
  \bibinfo {pages} {A17} (\bibinfo {year} {2024})},\ \Eprint
  {https://arxiv.org/abs/2208.01337} {arXiv:2208.01337 [astro-ph.HE]}
  \BibitemShut {NoStop}%
\bibitem [{\citenamefont {Moskalenko}\ \emph {et~al.}(2001)\citenamefont
  {Moskalenko}, \citenamefont {Mashnik},\ and\ \citenamefont
  {Strong}}]{Moskalenko:2001qm}%
  \BibitemOpen
  \bibfield  {author} {\bibinfo {author} {\bibfnamefont {I.~V.}\ \bibnamefont
  {Moskalenko}}, \bibinfo {author} {\bibfnamefont {S.~G.}\ \bibnamefont
  {Mashnik}},\ and\ \bibinfo {author} {\bibfnamefont {A.~W.}\ \bibnamefont
  {Strong}},\ }in\ \href@noop {} {\emph {\bibinfo {booktitle} {{27th
  International Cosmic Ray Conference}}}}\ (\bibinfo {year} {2001})\ \Eprint
  {https://arxiv.org/abs/astro-ph/0106502} {arXiv:astro-ph/0106502}
  \BibitemShut {NoStop}%
\bibitem [{\citenamefont {Moskalenko}\ and\ \citenamefont
  {Mashnik}(2003)}]{Moskalenko:2003kp}%
  \BibitemOpen
  \bibfield  {author} {\bibinfo {author} {\bibfnamefont {I.~V.}\ \bibnamefont
  {Moskalenko}}\ and\ \bibinfo {author} {\bibfnamefont {S.~G.}\ \bibnamefont
  {Mashnik}},\ }in\ \href@noop {} {\emph {\bibinfo {booktitle} {{28th
  International Cosmic Ray Conference}}}}\ (\bibinfo {year} {2003})\ pp.\
  \bibinfo {pages} {1969--1972},\ \Eprint
  {https://arxiv.org/abs/astro-ph/0306367} {arXiv:astro-ph/0306367}
  \BibitemShut {NoStop}%
\bibitem [{\citenamefont {Webber}\ \emph
  {et~al.}(1990{\natexlab{c}})\citenamefont {Webber}, \citenamefont {Kish},\
  and\ \citenamefont {Schrier}}]{Webber:1990kc}%
  \BibitemOpen
  \bibfield  {author} {\bibinfo {author} {\bibfnamefont {W.~R.}\ \bibnamefont
  {Webber}}, \bibinfo {author} {\bibfnamefont {J.~C.}\ \bibnamefont {Kish}},\
  and\ \bibinfo {author} {\bibfnamefont {D.~A.}\ \bibnamefont {Schrier}},\
  }\href {https://doi.org/10.1103/PhysRevC.41.547} {\bibfield  {journal}
  {\bibinfo  {journal} {Phys. Rev. C}\ }\textbf {\bibinfo {volume} {41}},\
  \bibinfo {pages} {547} (\bibinfo {year} {1990}{\natexlab{c}})}\BibitemShut
  {NoStop}%
\bibitem [{\citenamefont {Webber}\ \emph
  {et~al.}(1998{\natexlab{a}})\citenamefont {Webber}, \citenamefont {Soutoul},
  \citenamefont {Kish}, \citenamefont {Rockstroh}, \citenamefont {Cassagnou},
  \citenamefont {Legrain},\ and\ \citenamefont {Testard}}]{Webber:1998ex}%
  \BibitemOpen
  \bibfield  {author} {\bibinfo {author} {\bibfnamefont {W.~R.}\ \bibnamefont
  {Webber}}, \bibinfo {author} {\bibfnamefont {A.}~\bibnamefont {Soutoul}},
  \bibinfo {author} {\bibfnamefont {J.~C.}\ \bibnamefont {Kish}}, \bibinfo
  {author} {\bibfnamefont {J.~M.}\ \bibnamefont {Rockstroh}}, \bibinfo {author}
  {\bibfnamefont {Y.}~\bibnamefont {Cassagnou}}, \bibinfo {author}
  {\bibfnamefont {R.}~\bibnamefont {Legrain}},\ and\ \bibinfo {author}
  {\bibfnamefont {O.}~\bibnamefont {Testard}},\ }\href
  {https://doi.org/10.1103/PhysRevC.58.3539} {\bibfield  {journal} {\bibinfo
  {journal} {Phys. Rev. C}\ }\textbf {\bibinfo {volume} {58}},\ \bibinfo
  {pages} {3539} (\bibinfo {year} {1998}{\natexlab{a}})}\BibitemShut {NoStop}%
\bibitem [{\citenamefont {Webber}\ \emph {et~al.}(2003)\citenamefont {Webber},
  \citenamefont {Soutoul}, \citenamefont {Kish},\ and\ \citenamefont
  {Rockstroh}}]{Webber_2003}%
  \BibitemOpen
  \bibfield  {author} {\bibinfo {author} {\bibfnamefont {W.~R.}\ \bibnamefont
  {Webber}}, \bibinfo {author} {\bibfnamefont {A.}~\bibnamefont {Soutoul}},
  \bibinfo {author} {\bibfnamefont {J.~C.}\ \bibnamefont {Kish}},\ and\
  \bibinfo {author} {\bibfnamefont {J.~M.}\ \bibnamefont {Rockstroh}},\ }\href
  {https://doi.org/10.1086/344051} {\bibfield  {journal} {\bibinfo  {journal}
  {The Astrophysical Journal Supplement Series}\ }\textbf {\bibinfo {volume}
  {144}},\ \bibinfo {pages} {153} (\bibinfo {year} {2003})}\BibitemShut
  {NoStop}%
\bibitem [{\citenamefont {Otuka}\ \emph {et~al.}(2014)\citenamefont {Otuka}
  \emph {et~al.}}]{Otuka:2014wzu}%
  \BibitemOpen
  \bibfield  {author} {\bibinfo {author} {\bibfnamefont {N.}~\bibnamefont
  {Otuka}} \emph {et~al.},\ }\href {https://doi.org/10.1016/j.nds.2014.07.065}
  {\bibfield  {journal} {\bibinfo  {journal} {Nucl. Data Sheets}\ }\textbf
  {\bibinfo {volume} {120}},\ \bibinfo {pages} {272} (\bibinfo {year}
  {2014})},\ \Eprint {https://arxiv.org/abs/2002.07114} {arXiv:2002.07114
  [nucl-ex]} \BibitemShut {NoStop}%
\bibitem [{\citenamefont {Webber}\ \emph
  {et~al.}(1998{\natexlab{b}})\citenamefont {Webber}, \citenamefont {Kish},
  \citenamefont {Rockstroh}, \citenamefont {Cassagnou}, \citenamefont
  {Legrain}, \citenamefont {Soutoul}, \citenamefont {Testard},\ and\
  \citenamefont {Tull}}]{Webber_1998}%
  \BibitemOpen
  \bibfield  {author} {\bibinfo {author} {\bibfnamefont {W.~R.}\ \bibnamefont
  {Webber}}, \bibinfo {author} {\bibfnamefont {J.~C.}\ \bibnamefont {Kish}},
  \bibinfo {author} {\bibfnamefont {J.~M.}\ \bibnamefont {Rockstroh}}, \bibinfo
  {author} {\bibfnamefont {Y.}~\bibnamefont {Cassagnou}}, \bibinfo {author}
  {\bibfnamefont {R.}~\bibnamefont {Legrain}}, \bibinfo {author} {\bibfnamefont
  {A.}~\bibnamefont {Soutoul}}, \bibinfo {author} {\bibfnamefont
  {O.}~\bibnamefont {Testard}},\ and\ \bibinfo {author} {\bibfnamefont
  {C.}~\bibnamefont {Tull}},\ }\href {https://doi.org/10.1086/306446}
  {\bibfield  {journal} {\bibinfo  {journal} {The Astrophysical Journal}\
  }\textbf {\bibinfo {volume} {508}},\ \bibinfo {pages} {949} (\bibinfo {year}
  {1998}{\natexlab{b}})}\BibitemShut {NoStop}%
\bibitem [{\citenamefont {Reeder}(1965)}]{REEDER19651879}%
  \BibitemOpen
  \bibfield  {author} {\bibinfo {author} {\bibfnamefont {P.}~\bibnamefont
  {Reeder}},\ }\href
  {https://doi.org/https://doi.org/10.1016/0022-1902(65)80038-0} {\bibfield
  {journal} {\bibinfo  {journal} {Journal of Inorganic and Nuclear Chemistry}\
  }\textbf {\bibinfo {volume} {27}},\ \bibinfo {pages} {1879} (\bibinfo {year}
  {1965})}\BibitemShut {NoStop}%
\bibitem [{\citenamefont {Read}\ and\ \citenamefont
  {Viola}(1984)}]{READ1984359}%
  \BibitemOpen
  \bibfield  {author} {\bibinfo {author} {\bibfnamefont {S.}~\bibnamefont
  {Read}}\ and\ \bibinfo {author} {\bibfnamefont {V.}~\bibnamefont {Viola}},\
  }\href {https://doi.org/https://doi.org/10.1016/0092-640X(84)90009-3}
  {\bibfield  {journal} {\bibinfo  {journal} {Atomic Data and Nuclear Data
  Tables}\ }\textbf {\bibinfo {volume} {31}},\ \bibinfo {pages} {359} (\bibinfo
  {year} {1984})}\BibitemShut {NoStop}%
\bibitem [{\citenamefont {Napolitani}\ \emph {et~al.}(2004)\citenamefont
  {Napolitani}, \citenamefont {Schmidt}, \citenamefont {Botvina}, \citenamefont
  {Rejmund}, \citenamefont {Tassan-Got},\ and\ \citenamefont
  {Villagrasa}}]{Napolitani:2004fw}%
  \BibitemOpen
  \bibfield  {author} {\bibinfo {author} {\bibfnamefont {P.}~\bibnamefont
  {Napolitani}}, \bibinfo {author} {\bibfnamefont {K.~H.}\ \bibnamefont
  {Schmidt}}, \bibinfo {author} {\bibfnamefont {A.~S.}\ \bibnamefont
  {Botvina}}, \bibinfo {author} {\bibfnamefont {F.}~\bibnamefont {Rejmund}},
  \bibinfo {author} {\bibfnamefont {L.}~\bibnamefont {Tassan-Got}},\ and\
  \bibinfo {author} {\bibfnamefont {C.}~\bibnamefont {Villagrasa}},\ }\href
  {https://doi.org/10.1103/PhysRevC.70.054607} {\bibfield  {journal} {\bibinfo
  {journal} {Phys. Rev. C}\ }\textbf {\bibinfo {volume} {70}},\ \bibinfo
  {pages} {054607} (\bibinfo {year} {2004})},\ \Eprint
  {https://arxiv.org/abs/nucl-ex/0406006} {arXiv:nucl-ex/0406006} \BibitemShut
  {NoStop}%
\bibitem [{\citenamefont {Korejwo}\ \emph {et~al.}(2000)\citenamefont
  {Korejwo}, \citenamefont {Dzikowski}, \citenamefont {Giller}, \citenamefont
  {Wdowczyk}, \citenamefont {Perelygin},\ and\ \citenamefont
  {Zarubin}}]{Korejwo:2000pf}%
  \BibitemOpen
  \bibfield  {author} {\bibinfo {author} {\bibfnamefont {A.}~\bibnamefont
  {Korejwo}}, \bibinfo {author} {\bibfnamefont {T.}~\bibnamefont {Dzikowski}},
  \bibinfo {author} {\bibfnamefont {M.}~\bibnamefont {Giller}}, \bibinfo
  {author} {\bibfnamefont {J.}~\bibnamefont {Wdowczyk}}, \bibinfo {author}
  {\bibfnamefont {V.~V.}\ \bibnamefont {Perelygin}},\ and\ \bibinfo {author}
  {\bibfnamefont {A.~V.}\ \bibnamefont {Zarubin}},\ }\href
  {https://doi.org/10.1088/0954-3899/26/8/306} {\bibfield  {journal} {\bibinfo
  {journal} {J. Phys. G}\ }\textbf {\bibinfo {volume} {26}},\ \bibinfo {pages}
  {1171} (\bibinfo {year} {2000})}\BibitemShut {NoStop}%
\bibitem [{\citenamefont {Korejwo}\ \emph {et~al.}(2002)\citenamefont
  {Korejwo}, \citenamefont {Giller}, \citenamefont {Dzikowski}, \citenamefont
  {Perelygin},\ and\ \citenamefont {Zarubin}}]{Korejwo:2002ts}%
  \BibitemOpen
  \bibfield  {author} {\bibinfo {author} {\bibfnamefont {A.}~\bibnamefont
  {Korejwo}}, \bibinfo {author} {\bibfnamefont {M.}~\bibnamefont {Giller}},
  \bibinfo {author} {\bibfnamefont {T.}~\bibnamefont {Dzikowski}}, \bibinfo
  {author} {\bibfnamefont {V.~V.}\ \bibnamefont {Perelygin}},\ and\ \bibinfo
  {author} {\bibfnamefont {A.~V.}\ \bibnamefont {Zarubin}},\ }\href
  {https://doi.org/10.1088/0954-3899/28/6/304} {\bibfield  {journal} {\bibinfo
  {journal} {J. Phys. G}\ }\textbf {\bibinfo {volume} {28}},\ \bibinfo {pages}
  {1199} (\bibinfo {year} {2002})}\BibitemShut {NoStop}%
\bibitem [{\citenamefont {Bazarov}\ \emph {et~al.}(2005)\citenamefont
  {Bazarov}, \citenamefont {Glagolev}, \citenamefont {Lugovoi}, \citenamefont
  {Lutpullaev}, \citenamefont {Olimov}, \citenamefont {Petrov}, \citenamefont
  {Yuldashev},\ and\ \citenamefont {Yuldashev}}]{Bazarov:2005qu}%
  \BibitemOpen
  \bibfield  {author} {\bibinfo {author} {\bibfnamefont {E.~K.}\ \bibnamefont
  {Bazarov}}, \bibinfo {author} {\bibfnamefont {V.~V.}\ \bibnamefont
  {Glagolev}}, \bibinfo {author} {\bibfnamefont {V.~V.}\ \bibnamefont
  {Lugovoi}}, \bibinfo {author} {\bibfnamefont {S.~L.}\ \bibnamefont
  {Lutpullaev}}, \bibinfo {author} {\bibfnamefont {K.}~\bibnamefont {Olimov}},
  \bibinfo {author} {\bibfnamefont {V.~I.}\ \bibnamefont {Petrov}}, \bibinfo
  {author} {\bibfnamefont {A.~A.}\ \bibnamefont {Yuldashev}},\ and\ \bibinfo
  {author} {\bibfnamefont {B.~S.}\ \bibnamefont {Yuldashev}},\ }\href
  {https://doi.org/10.1134/1.1914868} {\bibfield  {journal} {\bibinfo
  {journal} {JETP Lett.}\ }\textbf {\bibinfo {volume} {81}},\ \bibinfo {pages}
  {140} (\bibinfo {year} {2005})}\BibitemShut {NoStop}%
\bibitem [{\citenamefont {{Yiou}}\ \emph {et~al.}(1969)\citenamefont {{Yiou}},
  \citenamefont {{Seide}},\ and\ \citenamefont
  {{Bernas}}}]{1969JGR....74.2447Y}%
  \BibitemOpen
  \bibfield  {author} {\bibinfo {author} {\bibfnamefont {F.}~\bibnamefont
  {{Yiou}}}, \bibinfo {author} {\bibfnamefont {C.}~\bibnamefont {{Seide}}},\
  and\ \bibinfo {author} {\bibfnamefont {R.}~\bibnamefont {{Bernas}}},\ }\href
  {https://doi.org/10.1029/JA074i009p02447} {\bibfield  {journal} {\bibinfo
  {journal} {Journal of Geophysical Research}\ }\textbf {\bibinfo {volume}
  {74}},\ \bibinfo {pages} {2447} (\bibinfo {year} {1969})}\BibitemShut
  {NoStop}%
\bibitem [{\citenamefont {Dostrovsky}\ \emph {et~al.}(1965)\citenamefont
  {Dostrovsky}, \citenamefont {Davis}, \citenamefont {Poskanzer},\ and\
  \citenamefont {Reeder}}]{Dostrovsky:1965zz}%
  \BibitemOpen
  \bibfield  {author} {\bibinfo {author} {\bibfnamefont {I.}~\bibnamefont
  {Dostrovsky}}, \bibinfo {author} {\bibfnamefont {R.}~\bibnamefont {Davis}},
  \bibinfo {author} {\bibfnamefont {A.~M.}\ \bibnamefont {Poskanzer}},\ and\
  \bibinfo {author} {\bibfnamefont {P.~L.}\ \bibnamefont {Reeder}},\ }\href
  {https://doi.org/10.1103/PhysRev.139.B1513} {\bibfield  {journal} {\bibinfo
  {journal} {Phys. Rev.}\ }\textbf {\bibinfo {volume} {139}},\ \bibinfo {pages}
  {B1513} (\bibinfo {year} {1965})}\BibitemShut {NoStop}%
\bibitem [{\citenamefont {Silberberg}\ and\ \citenamefont
  {Tsao}(1973{\natexlab{b}})}]{Silberberg:1973kxa}%
  \BibitemOpen
  \bibfield  {author} {\bibinfo {author} {\bibfnamefont {R.}~\bibnamefont
  {Silberberg}}\ and\ \bibinfo {author} {\bibfnamefont {C.~H.}\ \bibnamefont
  {Tsao}},\ }\href@noop {} {\bibfield  {journal} {\bibinfo  {journal}
  {Astrophys. J. Suppl.}\ }\textbf {\bibinfo {volume} {25}},\ \bibinfo {pages}
  {335} (\bibinfo {year} {1973}{\natexlab{b}})}\BibitemShut {NoStop}%
\bibitem [{\citenamefont {{Silberberg}}\ \emph {et~al.}(1985)\citenamefont
  {{Silberberg}}, \citenamefont {{Tsao}},\ and\ \citenamefont
  {{Letaw}}}]{1985ApJS...58..873S}%
  \BibitemOpen
  \bibfield  {author} {\bibinfo {author} {\bibfnamefont {R.}~\bibnamefont
  {{Silberberg}}}, \bibinfo {author} {\bibfnamefont {C.~H.}\ \bibnamefont
  {{Tsao}}},\ and\ \bibinfo {author} {\bibfnamefont {J.~R.}\ \bibnamefont
  {{Letaw}}},\ }\href {https://doi.org/10.1086/191058} {\bibfield  {journal}
  {\bibinfo  {journal} {The Astrophysical Journal Supplement Series}\ }\textbf
  {\bibinfo {volume} {58}},\ \bibinfo {pages} {873} (\bibinfo {year}
  {1985})}\BibitemShut {NoStop}%
\end{thebibliography}%

\appendix

\section{CROSS SECTION DATA\label{app:xsdata}}
In our previous work~\cite{Zhao:2024qbj}, we introduced data-driven parametrization by using the evaluation routine implemented in the {\footnotesize GALPROP} code.
This parametrization was constructed to make full use of the cross-section available data, which were obtained from:
\begin{enumerate}
    \item\texttt{isotope\_cs.dat}: The isotopic cross-section database file is built in the {\footnotesize GALPROP} code \cite{Moskalenko:2001qm,Moskalenko:2003kp} for normalizing the parametrization formulae, such as \texttt{WNEW} code by Webber \cite{Webber:1990kc,Webber:1998ex,Webber_2003} or \texttt{YIELDX} code by Tsao and Silberberg \cite{Silberberg:1998lxa}.
The cross-section data assembled in the file were taken from multiple cross-section measurements published before 2003.

\item EXFOR (Experimental Nuclear Reaction Data): The website\footnote{\url{https://www-nds.iaea.org/exfor}.} is an extensive database containing experimental data~\cite{Otuka:2014wzu}, as well as bibliographic information, experimental setup, and source of uncertainties. By querying the EXFOR database, we can add most of the measurements published so far.

\item NA61/SHINE: Additional measurements are reported by the NA61/SHINE Collaboration \cite{Amin:2021oow, Amin:2023fki}.
The recent pilot run provided precise high-energies measurements of cross sections from the C projectile at 13.5 GeV/$n$, which is valuable for constraining the uncertainties of the dominant channels.
\end{enumerate}
We have added thousands of data in \texttt{eval\_iso\_cs.dat} up to $\sim10$ GeVs to interpolate credible parametrization for important production channels of $\rm^2H$, $\rm^3He$, Li, Be, B, F, P, Sc, Ti and V.
In the published version of the paper, we will attach ancillary files in Supplemental Material that can be used to check the Li, B, and Be production cross-sections given in this work. These files (\texttt{eval\_iso\_cs.dat}, \texttt{isotope\_cs.dat} and \texttt{p\_cs\_fits.dat}) 
can be downloaded and added to the original files of the {\footnotesize GALPROP} code in the ``galtoolslib$\slash$nuclei'' folder for testing purposes.


Different from other measurements, the [We96] data~\cite{Ramaty:1996qt} are cumulative cross sections, hence the production of $\rm^6Li$ includes $\rm^6He$, and that of $\rm^7Li$ includes $\rm^7Be$. However, {\footnotesize GALPROP} incorrectly uses [We96] data as individual cross sections to renormalize the parametrization of [TS00]. According to the measurements from [We98], [Re65], and NUCLEX~\cite{Webber_1998, REEDER19651879, Genolini:2018ekk}, we estimate the cross sections of $\rm^{11}C$, $\rm^{9}Li$ and $\rm^{7}Be$. They are subtracted from the cumulative cross sections of [We96] to calculate the individual cross sections, and the derived data are listed in Table~\ref{tab:we96}.
The related cross-section data of $\rm^6He$ is not found, therefore we directly use the cumulative cross sections of $\rm^6Li+^6He$ to renormalize the parametrization.
For example, the cross section of $^{7}\text{Li}+p \longrightarrow ^{6}\text{Li}$ is 25.68 mb and the cross section of $^{7}\text{Li}+p \longrightarrow ^{6}\text{He}$ is 10.5 mb according to [TS00] parametrization above 2~GeV/$n$. We use [We96] data to renormalize the former cross section and the result is 31.48~mb above 2~GeV/$n$, which should be regarded as the cumulative cross section. Because of that, the cross section of $^{7}\text{Li}+p \longrightarrow ^{6}\text{He}$ should be set to zero to avoid multiple consideration of the ghost nuclei production.
The resulting fluxes of isotope $\rm^6Li$ and $\rm^7Li$ will both be reduced by 2\% at 10 GeV/$n$ due to the modification to [We96] data.

\begin{table}
\caption{\label{tab:we96} The cumulative cross-section data given in [We96]~\cite{Ramaty:1996qt} at 400~MeV/$n$ and the derived cross sections used in the work.}
\begin{ruledtabular}
\begin{tabular}{cccc}
 Channel& cumulative $\sigma$&derived $\sigma$&error\\ \hline
  $^{15}\text{N}+p \longrightarrow ^{11}\text{B}$&32.5 &26.9 &10\%\\
  $^{15}\text{N}+p \longrightarrow ^{9}\text{Be}$&7.5 &7.3 &20\%\\
  $^{15}\text{N}+p \longrightarrow ^{7}\text{Li}$&18.6 &12.92 &10\%\\
  $^{11}\text{B}+p \longrightarrow ^{9}\text{Be}$&14.6 &13.6 &30\%\\
  $^{11}\text{B}+p \longrightarrow ^{7}\text{Li}$&27.7 &24.5 &10\%\\
  $^{15}\text{N}+p \longrightarrow ^{6}\text{Li}$&10.5 &- &10\%\\
  $^{11}\text{B}+p \longrightarrow ^{6}\text{Li}$&6.3 &- &20\%\\
  $^{7}\text{Li}+p \longrightarrow ^{6}\text{Li}$&35.5 &- &10\%\\
\end{tabular}
\end{ruledtabular}
\end{table}

The measured cross section of $^{14}\text{N}+p \longrightarrow ^{6}\text{He}$ is 6.4~mb at 126 MeV/$n$ by [RV84]~\cite{READ1984359}. 
If we constantly extrapolate the [RV84] data to higher energies, the result would be one magnitude larger than the prediction of [TS00] parametrization, which is 0.78~mb above the plateau energy.
The reduction of cross sections at high energies also appears in the channels of $^{12}\text{C}+p \longrightarrow ^{6}\text{He}$ and $^{16}\text{O}+p \longrightarrow ^{6}\text{He}$, which could be a common feature.  Therefore, we assume that the cross section of $^{14}\text{N}+p \longrightarrow ^{6}\text{He}$ decreases exponentially when extrapolating to higher energies, which would finally reach 0.78~mb at 1~GeV/$n$.

\begin{figure}[htbp]
\includegraphics[width=0.5\textwidth,trim=0 0 0 0,clip]{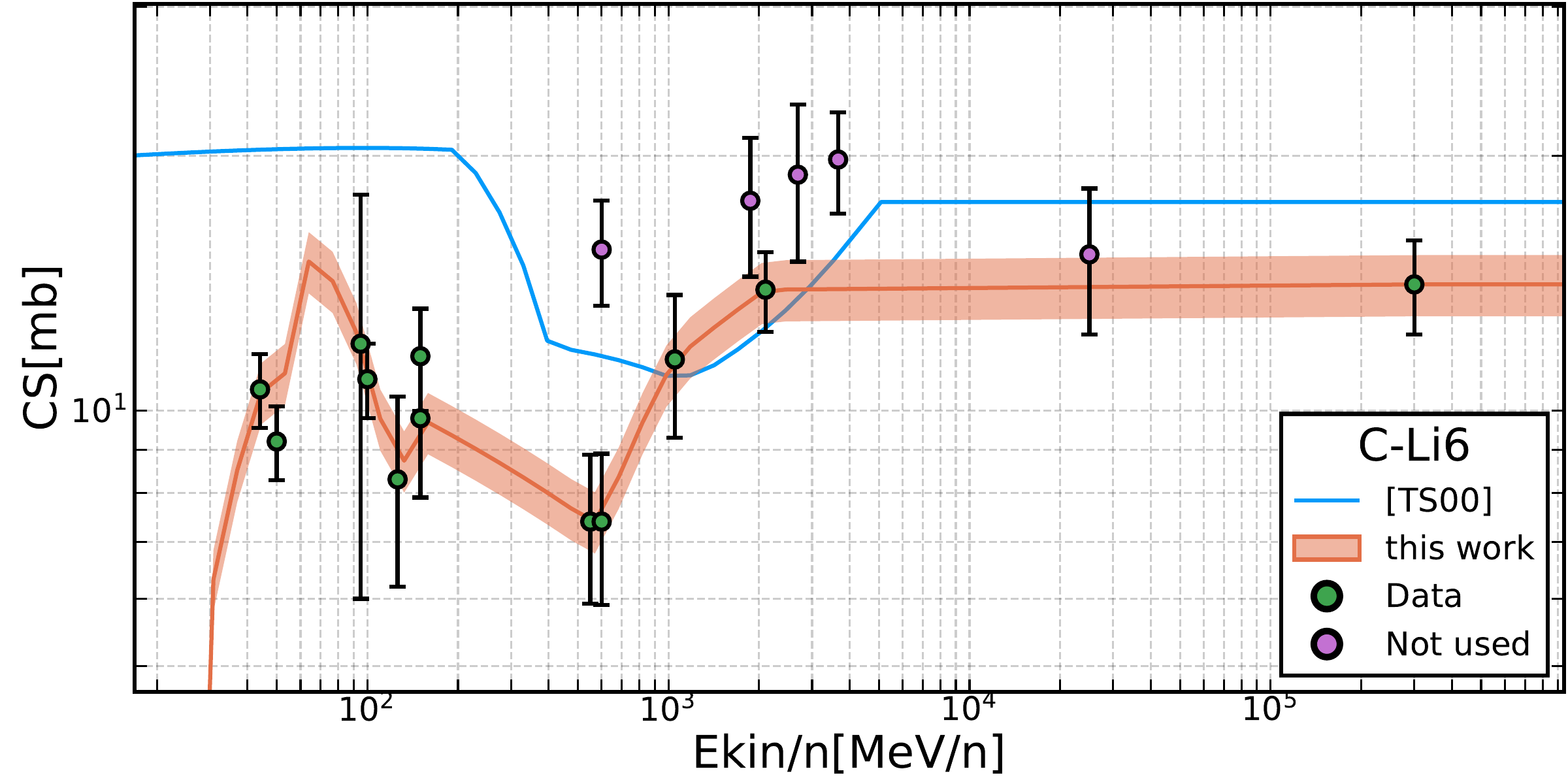}
\includegraphics[width=0.5\textwidth,trim=0 0 0 0,clip]{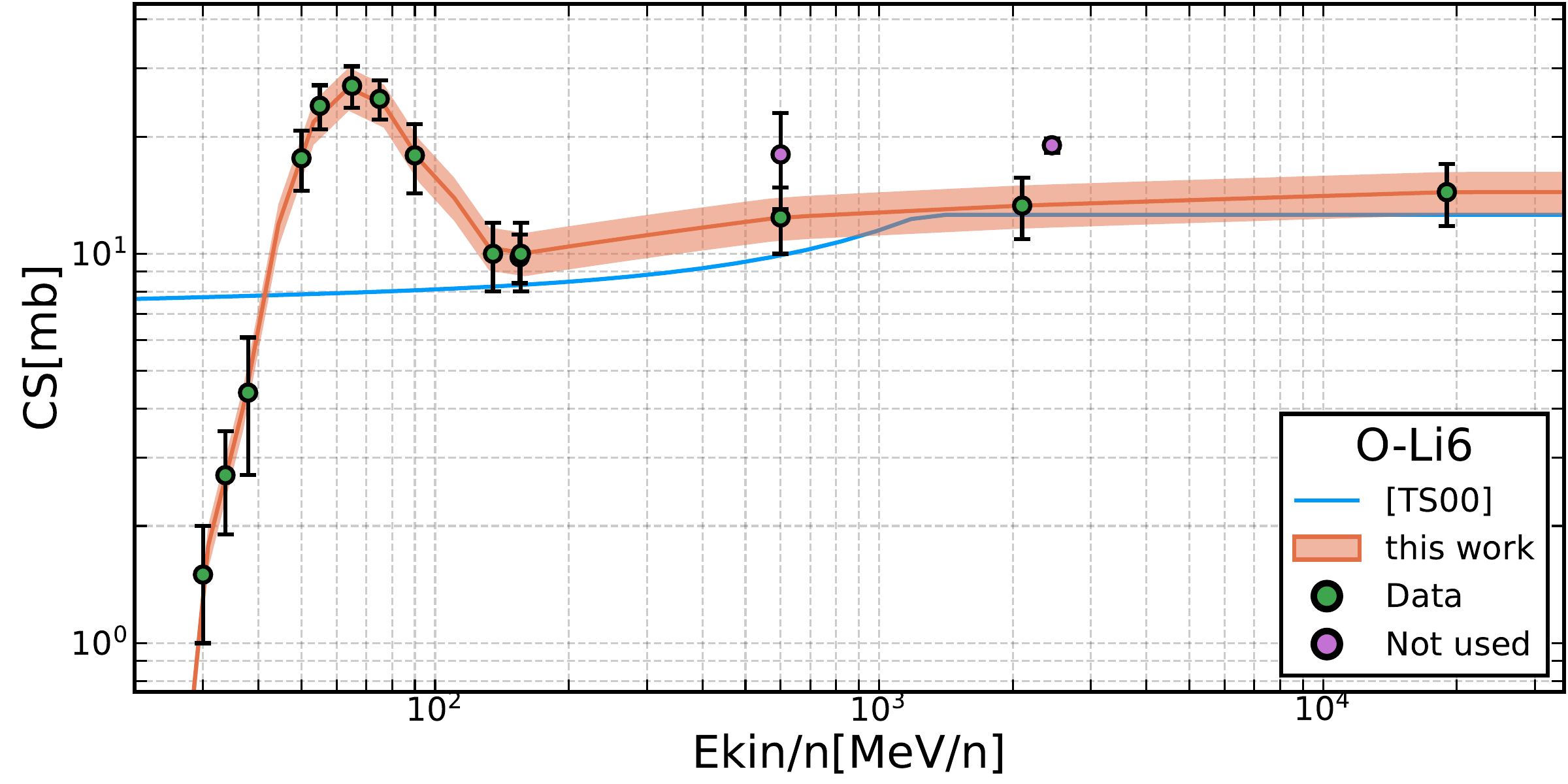}
\caption{\label{fig:xs_li} 
Cross-section measurements and the data-driven parametrization used in the work. The bands represent the cross-section uncertainties given in Table~\ref{tab:channel}. The parameterizations labeled as [TS00] were taken from the GALPROP code.
Top: Channel of $^{12}\text{C}+p \longrightarrow ^{6}\text{Li}$.
Bottom: Channel of $^{16}\text{O}+p \longrightarrow ^{6}\text{Li}$.
}
\end{figure}
\begin{figure}[htbp]
\includegraphics[width=0.5\textwidth,trim=0 0 0 0,clip]{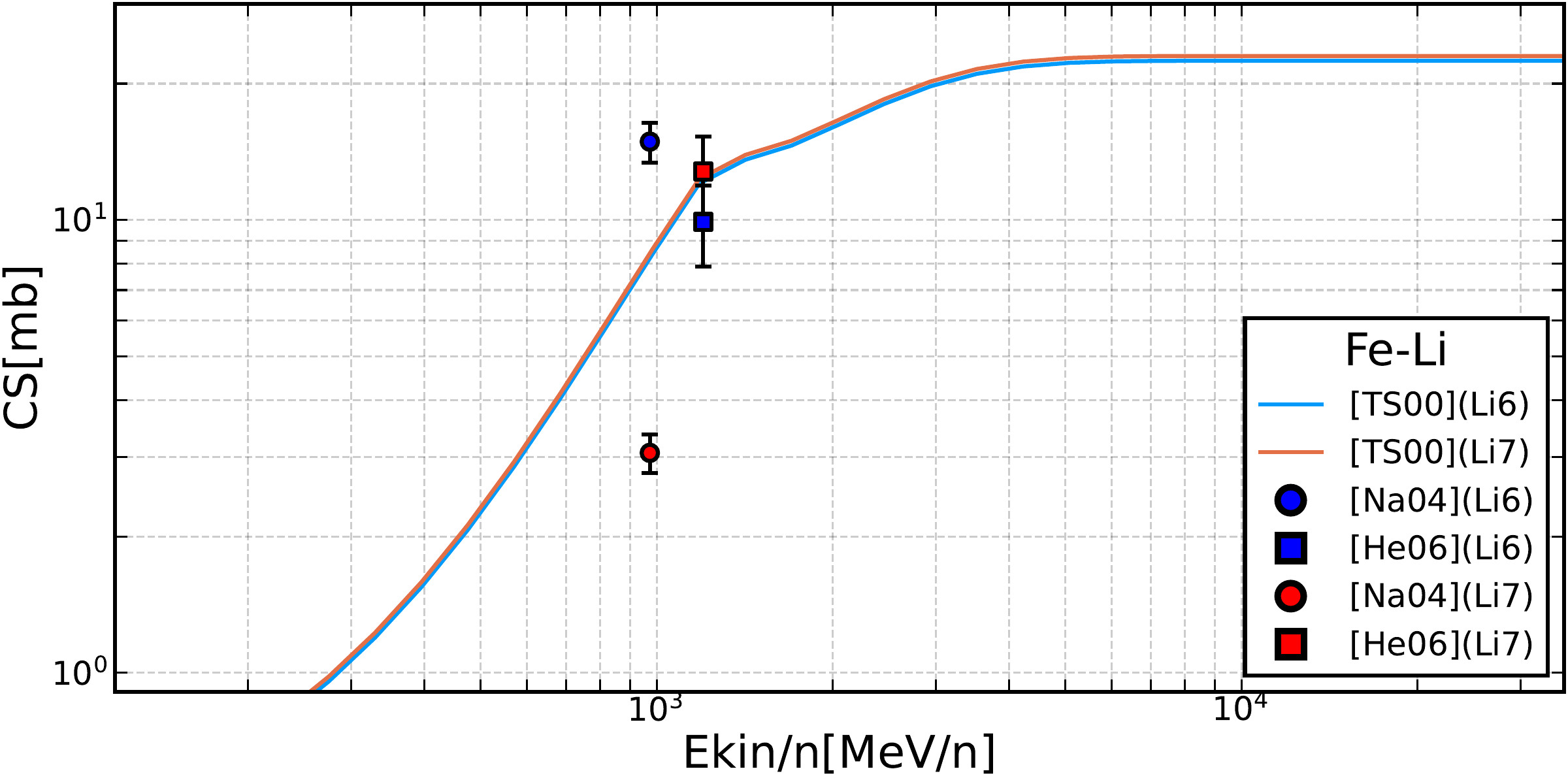}
\caption{\label{fig:xs_li2} 
Cross-section measurements of $^{56}\text{Fe}+p \longrightarrow ^{6}\text{Li}$ and $^{56}\text{Fe}+p \longrightarrow ^{7}\text{Li}$. The parameterizations labeled as [TS00] were taken from the GALPROP code. Data are from: [Na04]~\cite{Napolitani:2004fw} and [He06]~\cite{Herbach:2006rw}.
}
\end{figure}
As shown in Figure~\ref{fig:xs_li}, the data-driven cross sections of $^{12}\text{C}+p \longrightarrow ^{6}\text{Li}$ and $^{16}\text{O}+p \longrightarrow ^{6}\text{Li}$ used in the work are determined by the available measurements.
Considering that the cross section would not change rapidly above the plateau energy, we discarded some high-energy data (purple points) which are significantly larger.
These group's data ([Ko99], [Ko02], and [Ba05]~\cite{Korejwo:2000pf, Korejwo:2002ts, Bazarov:2005qu}) are also disfavored by our previous estimation of Be isotopes~\cite{Zhao:2024qbj}, which await to be verified by more high-energy measurements.
The parametrization results of [TS00] (blue lines) are shown, which fit the data-driven expectation of $^{16}\text{O}+p \longrightarrow ^{6}\text{Li}$, but the $^{12}\text{C}+p \longrightarrow ^{6}\text{Li}$ result of [TS00] is larger.
According to the $\rm^6Li$ result shown in Sec.~\ref{sec:result}, the discarded data and the parametrization results of [TS00] should be disfavored by the AMS-02 observations.

Besides C and O, the fragmentation of Fe is also important. As shown in Figure~\ref{fig:xs_li2}, the available measurements are from [Na04]~\cite{Napolitani:2004fw} and [He06]~\cite{Herbach:2006rw} respectively at 973 MeV/$n$ and 1200 MeV/$n$.
The [Na04] cross section of $\rm^6Li$ is about one magnitude larger than that of $\rm^7Li$, which contradicts the result of [He06] data as well as the [TS00] expectation.
Therefore, we only use the data of [He06] to renormalize the cross sections from Fe progenitor.

\section{RESCALE the RESULT of [TS00] PARAMETRIZATION\label{app:renorm}}
Webber's parametrization~\cite{Webber:1990kc, Webber:1998ex, Webber_2003} doesn't define the secondary Li production, and only the contributions of the dominant channels are introduced in the file \texttt{eval\_iso\_cs.dat} of {\footnotesize GALPROP} code for calculating the Li flux. 
To improve that, we use the [TS00] parametrization~\cite{Silberberg:1998lxa} instead for specific channels, when Webber's parametrization cannot calculate a non-zero cross-section result.

For the production of $\rm^6Li$, $\rm^7Li$ and $\rm^6He$ from projectiles heavier than O, the parametrization of [TS00] are obtained by an interpolation of measurements from $\rm^{12}C$ and $\rm^{16}O$~\cite{1969JGR....74.2447Y} and the derived result by the Monte Carol technique~\cite{Dostrovsky:1965zz} on the heavier projectiles (Z$>$28).
For less-measured reactions, the cross sections of [TS00] are renormalized to the available data to improve reliability. For none-measured reactions, the cross sections are used directly, which could be unreliable.
Li production from secondary CRs such as Be and B also relies seriously on extrapolating available measurements to lighter projectiles.
Those would lead to systematical overestimation or underestimation of cross sections from unconstrained channels and impact the Li spectrum at all energies.

Using the \texttt{YIELDX} code ([TS00]) implemented in {\footnotesize GALPROP}, we calculate the plateau-energy (high enough to be energy-independent) cross sections of $\rm^6Li$, $\rm^7Li$, and $\rm^6He$ as a function of the progenitor mass, compared with available high-energy measurements mentioned in Appendix.~\ref{app:xsdata}.
The [He06] data~\cite{Herbach:2006rw} are included to constrain the cross sections of heavy projectiles, as they measured the cross sections from projectiles heavier than Al at 1.2~GeV/$n$.
However, 1.2~GeV/$n$ is not high enough for the cross sections to reach the plateau and cannot be directly compared with [TS00].
Here we adopted the energy dependencies given by [TS00]~\cite{Silberberg:1973jxa, Silberberg:1973kxa} to scale the [He06] data from 1.2~GeV/$n$ to plateau energies.

As shown in Figure~\ref{fig:xsscale_li}, the cross sections of $\rm^6Li$ and $\rm^6He$ from Al to Ni ($27\leq A\leq58$) are overestimated by [TS00], while that of $\rm^7Li$ fits well with the data.
Therefore, we assume that the $\rm^6Li$ and $\rm^6He$ fragments from unmeasured reactions such as channels from Ne, Mg, and Si are also overestimated and should be reduced together.
The renormalize factors of $\rm^6Li$ and $\rm^6He$ are set to 0.8 and 0.7 respectively, which are roughly determined by fitting the data.
For the [He06] data heavier than Ni, the cross sections rapidly increase with the projectile mass, but the [TS00] parametrization fails to fit the data, except for the renormalized cross sections of $\rm^6He$ (green dot).
The result implies that the enhancement factors (introduced in Ref.~\cite{1985ApJS...58..873S}) for $\rm^6He$ may be correct, but enhancement factors for $\rm^6Li$ and $\rm^7Li$ should be re-evaluated according to the [He06] data.
However, this is beyond our studies, as the nuclei heavier than Ni are less abundant and, therefore not considered in the CR propagation and the nuclear reaction network.

In the top panel of Figure~\ref{fig:xsscale_li}, the cross sections of $^{9}\text{Be}+p \longrightarrow ^{6}\text{Li}$ and $^{11}\text{B}+p \longrightarrow ^{6}\text{Li}$ are illustrated, and we notice that [TS00]'s result are significantly larger than these data.
It could be possible that the cross sections of $\rm^6Li$ from most of the projectiles smaller than $\rm^{12}C$ are overestimated and should be reduced.
Here, we assume that the renormalize factor of $\rm^6Li$ for projectiles smaller than C is 0.33, which is determined by fitting the available data.
In the center panel, the [TS00]'s result of $^{11}\text{B}+p \longrightarrow ^{7}\text{Li}$ undershoots the data, and the result of$^{9}\text{Be}+p \longrightarrow ^{7}\text{Li}$ overshoots the data.
Therefore we cannot find a common renormalization for the cross sections of $\rm^7Li$ from light projectiles.
In the bottom panel, there is no available measurement for the cross sections of $\rm^6He$ from light projectiles. Considering that the renormalized parametrization (green dot) is globally consistent with the data, we utilize the renormalize factor of 0.7 for all the projectiles.

It is important to note that the cross-section features of $^{7}\text{Li}+p \longrightarrow ^{6}\text{Li}$ and $^{7}\text{Li}+p \longrightarrow ^{6}\text{He}$ are different from others.
In [TS00] parametrization~\cite{Silberberg:1973kxa}, these channels are regarded as peripheral reactions (the projectile only lose a proton or a neutron), whose formulae are independently defined and usually have larger cross sections. This result should not be useful for commonly constraining the cross sections of light projectiles, hence is excluded when calculating the renormalize factor.
 
Consequently, we find that the production cross sections of $\rm^6Li$ should be scaled by 0.8 for projectiles heavier than O, and should be scaled by 0.33 for projectiles lighter than C, the production cross sections of $\rm^6He$ should be scaled by 0.7 for all the unmeasured reactions.
These modifications are used in Sec.\ref{sec:xs}, and the impact to $\rm^6Li$ flux is illustrated in Figures~\ref{fig:isoLI_flux} and \ref{fig:li6change}.

\end{document}